\documentclass[twocolumn,aps,prb,floatfix]{revtex4}
\usepackage{graphics}
\usepackage{subfigure}
\usepackage{epsfig}

\makeatletter

\usepackage{verbatim}

\makeatletter

\input epsf

\def\be{\begin{equation}}
\def\ee{\end{equation}}
\def\ba{\begin{eqnarray}}
\def\ea{\end{eqnarray}}

\makeatother

\begin{document}

\title{Spin Waves in Striped Phases}

\author{E.~W.~Carlson$^1$, D.~X.~Yao$^2$, and D.~K.~Campbell$^2$}

\affiliation{(1) Dept. of Physics, Purdue University, West Lafayette, IN  47907 \\
(2) Dept. of Physics and Dept. of Electrical and Computer Engineering,
Boston University, Boston, MA 02215}

\date{\today}

\begin{abstract}

In many antiferromagnetic, quasi-two-dimensional materials, doping with holes
leads to ``stripe'' phases, in which the holes 
congregate along antiphase domain walls in the
otherwise antiferromagnetic texture.  Using a suitably parametrized
two-dimensional Heisenberg model on a square lattice, we study the spin wave spectra 
of well-ordered spin stripes, 
comparing bond-centered antiphase domain walls to 
site-centered antiphase domain walls for a range of spacings between the stripes and for stripes
both aligned with the lattice (``vertical'') and oriented along the diagonals of the lattice (``diagonal'').  
Our results establish that there are qualitative differences between the expected neutron
scattering responses 
for the bond-centered and site-centered cases. In particular, bond-centered stripes
of odd spacing generically exhibit more elastic peaks than their site-centered counterparts.  
For inelastic scattering, we find that 
bond-centered stripes produce more spin wave bands than site-centered stripes of 
the same spacing and that
bond-centered stripes produce rather isotropic low energy
spin wave cones for a large range of parameters, despite local microscopic anisotropy.  
We find that extra scattering intensity due to 
the crossing of spin wave modes (which may be linked to the ``resonance peak'' in the cuprates) 
is more likely for diagonal stripes, whether site- or bond-centered, whereas 
spin wave bands generically repel, rather than cross,  when stripes are vertical.
\end{abstract}
\maketitle

\section{Introduction}

Many doped strongly correlated materials exhibit evidence for an emergent length
scale in the form of ``stripes'', i.e., regular antpihase domain walls in an otherwise
antiferromagnetic texture.  
The strongest evidence for striped structures in nickelate perovskites and
some related cuprates
has come from neutron
scattering,\cite{tranquadareview,ybco,john,boothroyd,boothroyd03a} 
which is capable of detecting the spin texture
directly through diffraction.  Since several theories of high temperature 
superconductivity make contact with such
structures,\cite{spingap,zaanennature,sachdev,whitepair2001,markiewicz,recoil,marchida01b}
it is important to improve our 
microscopic picture of them. In particular, it is not yet known from experiment
whether the antiphase domain walls
sit primarily on nickel (copper) sites, 
or rather sit primarily on oxygen sites.

\begin{figure}[Htb]
{\centering
  \subfigure[Vertical site-centered]{\resizebox*{0.4\columnwidth}{!}{\includegraphics{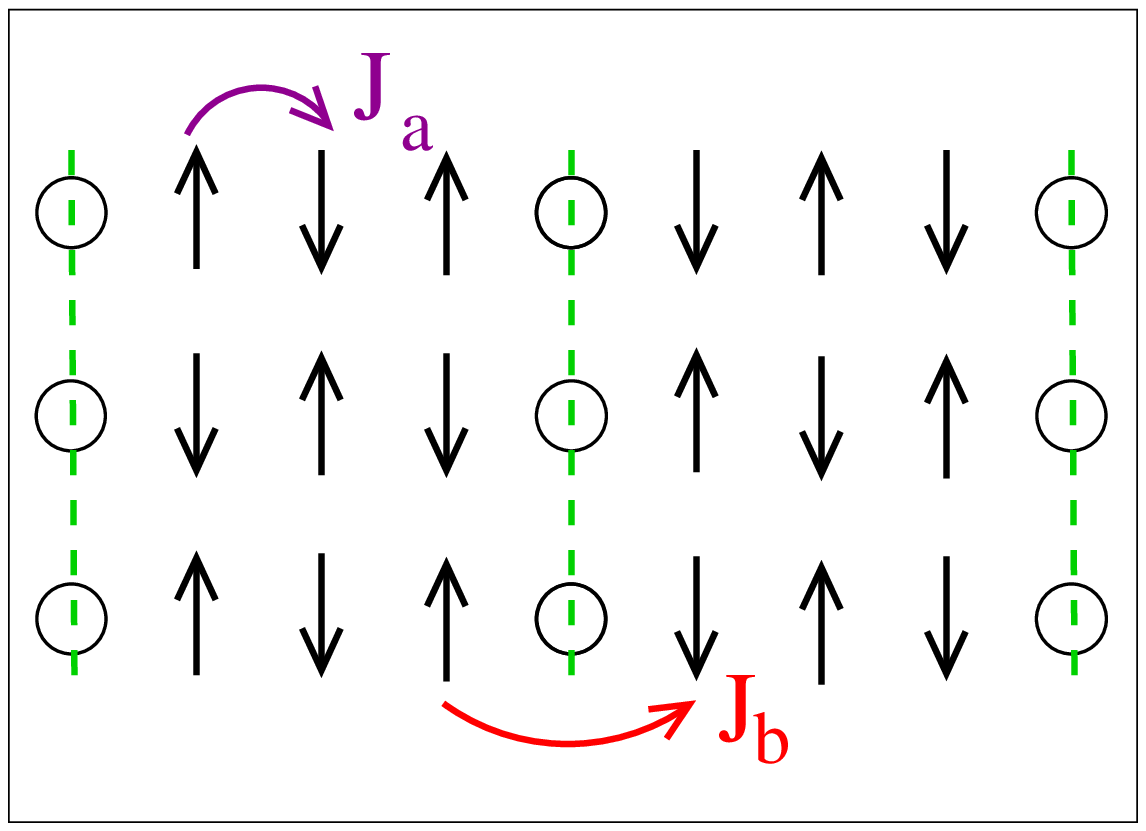}}}
  \subfigure[Vertical bond-centered]{\resizebox*{0.4\columnwidth}{!}{\includegraphics{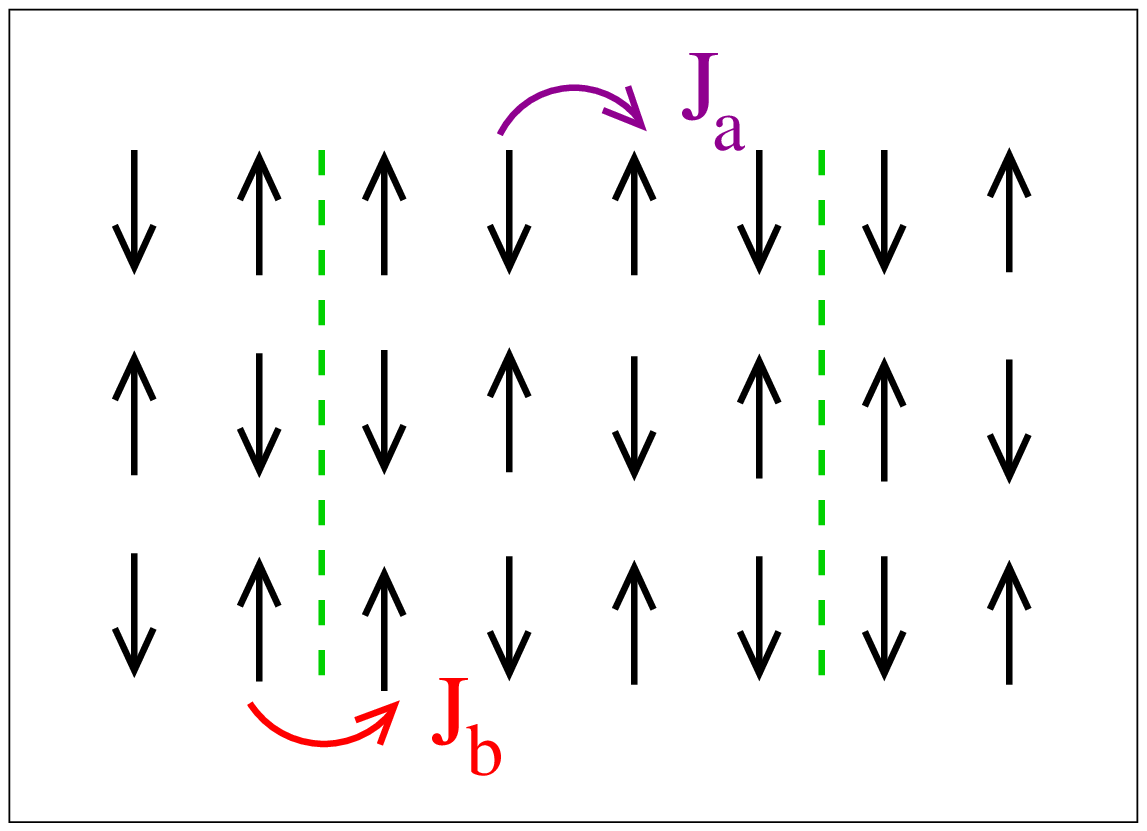}}} \par}
{\centering  \subfigure[Diagonal site-centered]{\resizebox*{0.4\columnwidth}{!}{\includegraphics{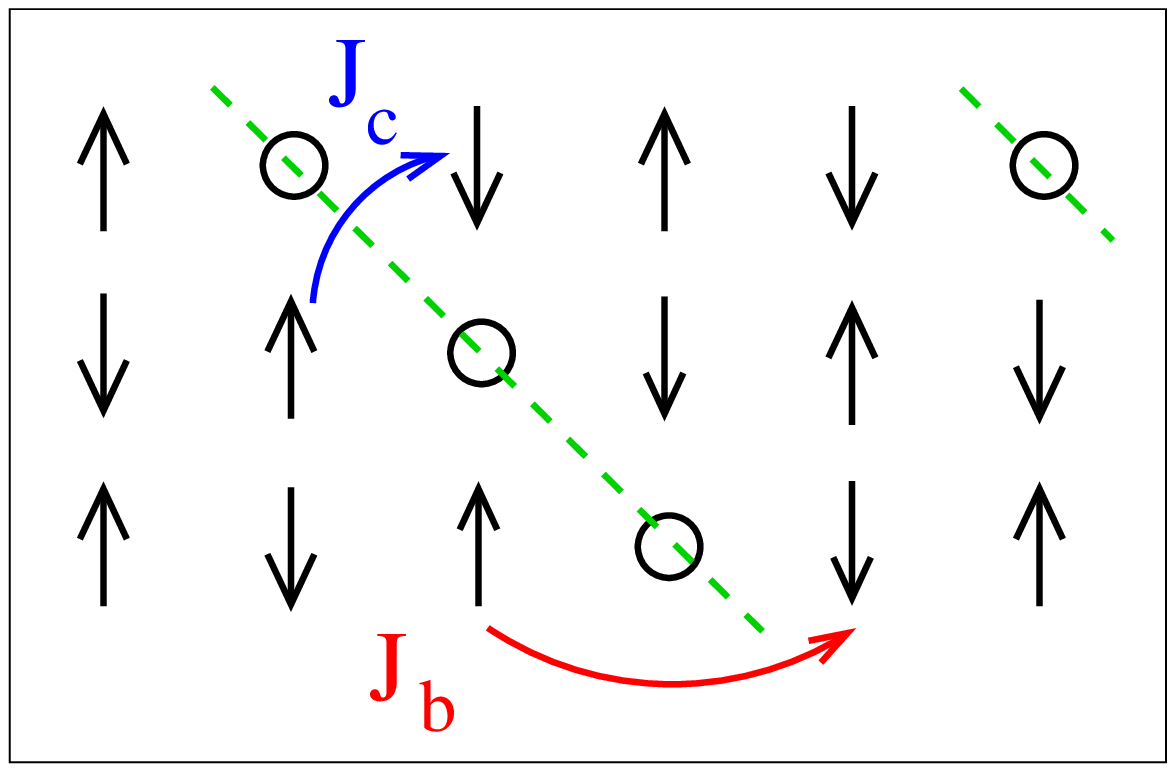}}}
  \subfigure[Diagonal bond-centered]{\resizebox*{0.4\columnwidth}{!}{\includegraphics{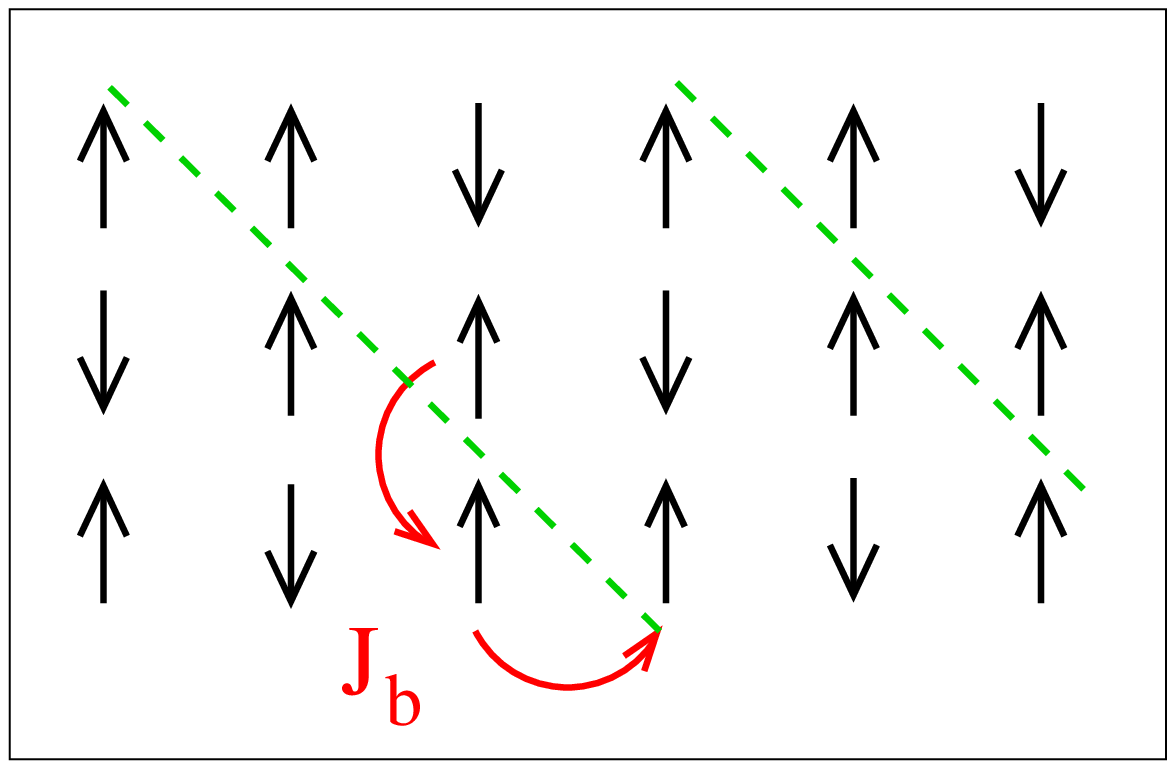}}}   \par}
\caption{(Color online) (a) Site-centered vertical stripe pattern with $p=4~$  lattice constants between domain walls.
In this configuration, exchange couplings $J_a>0$ and
  $J_b>0$ are all antiferromagnetic. (b) Bond-centered vertical stripe pattern with spacing $p=4$. The exchange
  coupling $J_a>0$ is antiferromagnetic, while $J_b<0$ is ferromagnetic.
(c) Diagonal site-centered domain walls
have coupling $J_b>0$ for next nearest neighbor spins coupled across the domain wall 
along the vectors $(2,0)$ and $(0,2)$, and coupling $J_c>0$ diagonally
to ``Manhattan'' second neighbors across the domain walls along the vector
$(1,1)$, in units where the square lattice spacing $a=1$.  (d) Diagonal bond-centered domain walls have nearest 
neighbor ferromagnetic coupling $J_b<0$ 
across the domain wall.  The size of each figure has been chosen for visual clarity.
\label{couplings}}
\end{figure}

When undoped, the nickel-oxygen (and copper-oxygen) planes in these materials are 
antiferromagnetic, with spin moments localized on the Ni (Cu) sites,
as evidenced by a peak in elastic neutron scattering 
at $(\pi,\pi)$.\cite{foot1}
Upon hole doping, this peak is observed to split into
four (or perhaps two\cite{ybco1d,wakimoto})
``incommensurate peaks'',\cite{foot2}
indicating an
extra modulation on top of the antiferromagnetic wavelength.
For the case of colinear spins, this is consistent with the formation of periodic
antiphase domain walls in the antiferromagnetic texture ({\em i.e.} stripes).

On a two-dimensional square lattice, 
these domains consist of a strip of antiferromagnet with spin up on, say, 
the ``A'' sublattice, separated by a domain wall from a strip of
antiferromagnet with spin up on the ``B'' sublattice, and so on,
as in Fig.~\ref{couplings}.  
The figures necessarily depict a certain
width for each antiphase domain wall, but the widths are not known and
are in reality likely less sharp than shown in the figure. In both cases,
neighboring antiferromagnetic patches have spin up on opposite 
sublattices, which washes out any signal at the antiferromagnetic peak $(\pi,\pi)$.  
Rather, satellite peaks are observed around $(\pi,\pi)$, at a distance determined
by the spacing between domain walls.
When the domain walls are site-centered, all couplings are antiferromagnetic,
including couplings across the domain walls.
Bond-centered domain walls, however, have some ferromagnetic couplings.\cite{sitecentered}
That is, bond-centered configurations consist of antiferromagnetic patches 
which are {\em ferromagnetically} coupled across the domain wall.  
As shown in Fig.~\ref{couplings}, we consider stripes aligned with the lattice direction (called 
``vertical stripes'') or aligned along the lattice diagonals (called ``diagonal stripes'').

In this article we focus on the spin wave spectra and 
expected magnetic scattering intensities of bond-centered
and site-centered stripe phases of various spacings and orientations. 
Other stripe phases are certainly possible,
such as phases which mix site- and bond-centered domain walls, or phases
in which the spacing of the antiphase domain walls is not commensurate 
with the underlying
lattice, or ``dynamic'' stripes\cite{stripefluct}, which fluctuate in time,
We will not consider these cases here, but focus on well-ordered
spin stripes which have purely site- or bond-centered domain walls.  As we will show below,
there are qualitative differences between the spin wave spectra of 
bond- and site-centered domain walls, indicating that in some cases inelastic neutron
scattering may be able to distinguish between the two.  In addition,
there is a difference in the number of peaks in the elastic spin
structure factor for odd stripe spacings, indicating that elastic neutron scattering alone
may be able to distinguish as well.  

\section{Model}

We consider static, ordered arrays of antiphase domain walls in an
otherwise antiferromagnetic texture.  Although the domain
walls collect charge\cite{zkelandau,zaanenfluid,whitestripesallx2d,ichikawa}, we neglect 
this charge component,
as we are interested solely in the response of the spin degrees of freedom.
We use a Heisenberg model on a two-dimensional square lattice:
\begin{equation}
H= \frac{1}{2} \sum_{<\mathbf{r},\mathbf{r'}>} J_{\mathbf{r},\mathbf{r'}} \mathbf{S}_{\mathbf{r}} \mathbf{S}_{\mathbf{r'}},
\label{model}
\end{equation}
where \(<\mathbf{r},\mathbf{r'}>\) runs over all spin sites, and the exchange
coupling is \(J_{\mathbf{r},\mathbf{r'}}\).  Within an antiferromagnetic
patch, nearest neighbor couplings
are antiferromagnetic with $J_{\mathbf{r},\mathbf{r'}}=J_a>0$.   
Couplings across a domain wall depend upon the configuration and are enumerated below.  
All other couplings are neglected.
When comparing to the nickel oxides (copper oxides), our lattice corresponds to 
the nickel (copper) sites within the nickel-oxygen (copper-oxygen) planes.

\subsection{Vertical Stripes}
We consider first the case where stripes run parallel to the Ni-O (Cu-O) bond direction;
we call these ``vertical'' stripes.
As illustrated in Fig.~\ref{couplings},
when the domain wall is centered
on a lattice site, we may describe the system as having no net spin on the
domain wall.\cite{foot3}
In this case, spins from the edges of neighboring antiferromagnetic
patches are 
coupled across the domain wall {\em antiferromagnetically},
$J_{\mathbf{r},\mathbf{r'}}=J_b>0$
 with $\mathbf{S}_{\mathbf{r}}=0$
on the domain wall, as illustrated in Fig.~\ref{couplings}(a).  
Within the antiferromagnetic patches, nearest neighbor spins are of 
course also antiferromagnetically coupled, $J_{\mathbf{r},\mathbf{r'}}=J_a>0$.
When, however, the domain wall is bond-centered---that is, situated between two sites
as in Fig.~\ref{couplings}(b)--- 
spins from the edges of neighboring antiferromagnetic patches are 
{\em ferromagnetically} coupled, and we have
$J_{\mathbf{r},\mathbf{r'}}=J_{b}<0$ across the domain wall.
Nearest neighbor exchange couplings within each antiferromagnetic patch remain
antiferromagnetic, $J_{\mathbf{r},\mathbf{r'}}=J_a>0$.  
We shall see that this ferromagnetic coupling $J_b$ of spins across the domain wall leads to distinctive
features for the spin waves in the bond-centered case.

We define the magnetic Bravais lattice as follows.\cite{kruger}
Let $p$ denote the distance between domain walls.
We will henceforth work in units where the square lattice spacing $a=1$.
For $p = odd$,  we choose the 
basis vectors $\mathbf{A}_1=(p,0)$ and $\mathbf{A}_2=(0,2)$, and
for $p = even$, we use 
$\mathbf{A_1}=(p,1)$ and $\mathbf{A}_2=(0,2)$.
For site-centered configurations, 
there are $N=2p$ sites within each unit cell which include
$2(p-1)$ spins and 2 sites with no static spin component. 
For bond-centered domain walls, there are $N=2p$ spins in each unit cell. 
(See Fig.~\ref{vertical}.)

We use the notation $VSp$ and $VBp$ to refer to vertical stripes of spacing $p$ in a 
site($S$)- or bond($B$)-centered configuration, respectively.  For example,
$VS3$ refers to a vertical site-centered configuration with spacing $p=3$
between domain walls.

\subsection{Diagonal Stripes \label{sec:diag}}
For diagonal stripes, the antiphase domain walls are oriented along the
$(1, \pm 1)$ direction in a square lattice (recall we have set the lattice
spacing $a=1$).   
For the same microscopic interaction strengths
(deriving $J_{\mathbf{r},\mathbf{r'}}$ from, {\em e.g.}, a Hubbard model), spins are more strongly coupled across
the domain wall than in the vertical case.  For example, with diagonal bond-centered stripes,
each spin neighboring the domain wall interacts with {\em two} nearest neighbor 
(ferromagnetically coupled)
spins across the domain wall, as shown in Fig.~\ref{couplings}(d).  
Contrast this with the vertical stripes of Fig.~\ref{couplings}(a) and (b),
where each spin neighboring a domain wall interacts with only one spin across the domain wall.
Diagonal site-centered
stripes are even more strongly coupled, with two different types of interactions across the
domain wall, one of which we label $J_b$ because it connects spins
along a bond direction (connecting spins along the vectors  
$(2,0)$
and $(0,2)$
across the
domain wall), and the other we label $J_c$ (connecting spins along the vector 
$(1,1)$ across the domain wall),
as shown in Fig.~\ref{couplings}(c).

For diagonal stripes, the magnetic Bravais lattice differs from the vertical case. 
For $p= odd$ spacing between domain walls,  we choose the 
basis vectors $\mathbf{A}_1=(p,0)$ and $\mathbf{A}_2=(-1,1)$, and
for $p= even$, we use 
$\mathbf{A_1}=(2p,0)$ and $\mathbf{A}_2=(-1,1)$.
For site-centered configurations, 
when $p$ is even 
there are $N=2p$ sites within each unit cell which includes
$2(p-1)$ spins and 2 sites with no static spin component,
and when $p$ is odd, there are $N=p$ sites within each unit cell, which includes
$p-1$ spins, and one empty site. 
For bond-centered domain walls, there are $N=2p$ spins in each unit cell when
$p$ is even, and there are $N=p$ spins in the unit cell when $p$ is odd. 
(See Fig.~\ref{diagonal}.)

We use the notation $DSp$ and $DBp$ to refer to diagonal stripes of spacing $p$ in a 
site ($S$)- or bond ($B$)-centered configuration, respectively.  

\section{Spin Wave Theory \label{theory}}

The elementary excitations of ordered spin textures may be studied
using the well-known technique of Holstein-Primakoff bosons.
The same dispersion is obtained by quantizing the classical spin waves,
and the methods are equivalent as $S\rightarrow \infty$.  
We use each description when convenient. 
As it is physically more transparent, we review here the latter method\cite{goodstein},
discussing the former in Appendix~\ref{hp}.

In the classical spin wave approach, each spin is treated as precessing 
in the effective field produced by its coupled neighbors, via
the torque equations of a spin in a magnetic field.\cite{goodstein}
The rate of change of the spin at position \(\mathbf{r}\) is described by
\begin{equation}
\hbar \frac{d \mathbf{S}_{\mathbf{r}}}{d t}= \mathbf{\mu}_{\mathbf{r}} \times
\mathbf{H}_{\mathbf{r}}^{eff}~, \label{torque}
\end{equation}
where $\mathbf{\mu}_{\mathbf{r}}$ and $\mathbf{H}_{\mathbf{r}}^{eff}$ are respectively 
the corresponding
magnetic moment and effective magnetic field at position \(\mathbf{r}\), defined by
\begin{eqnarray} 
\mathbf{\mu}_{\mathbf{r}} &=& -g \mu_B
  \mathbf{S}_{\mathbf{r}}  \nonumber \\
\mathbf{H}_{\mathbf{r}}^{eff} &=& \frac{1}{g \mu_{B}} \sum_{\mathbf{r'}}
J_{\mathbf{r},\mathbf{r'}} \mathbf{S}_{\mathbf{r'}}~,  
\end{eqnarray}

Within our model, Eqn.~(\ref{model}), the torque equations become
\begin{eqnarray}
\frac{d S^x_{\mathbf{r}}}{d t}= && -\frac{1}{\hbar} (
S^y_{\mathbf{r}} \sum_{\mathbf{r'}}
J_{\mathbf{r},\mathbf{r'}} S^z_{\mathbf{r'}}-S^z_{\mathbf{r}} \sum_{\mathbf{r'} }
J_{\mathbf{r},\mathbf{r'}} S^y_{\mathbf{r'}} )   \nonumber \\
\frac{d S^y_{\mathbf{r}}}{d t}= && -\frac{1}{\hbar} (S^z_{\mathbf{r}}
\sum_{\mathbf{r'} } J_{\mathbf{r},\mathbf{r'}}
S^x_{\mathbf{r'}}-S^x_{\mathbf{r}} \sum_{\mathbf{r'}}
J_{\mathbf{r},\mathbf{r'}} S^z_{\mathbf{r'}} )   \nonumber \\ 
\frac{d S^z_{\mathbf{r}}}{dt} \approx &&0~,
\end{eqnarray}
where we have assumed large $S$ and small oscillations, so that changes
in $S^z$ can be neglected.
We seek solutions of the form
\begin{eqnarray}
S_{\mathbf{r}}^x=S_i^x \exp{[i(\mathbf{k} \cdot \mathbf{r} -\omega t)]} \nonumber \\
S_{\mathbf{r}}^y=S_i^y \exp{[i(\mathbf{k} \cdot \mathbf{r} -\omega t)]}
\end{eqnarray}
where i labels spins within the unit cell, i.e. $i=1, 2, \cdots , N $; N is
the total number of spins in the unit cell; $\mathbf{k}=(k_x, k_y)$, and $\mathbf{r}=(r_x, r_y)$.
Setting the determinant of the coefficients of
$S_i^x$ and $S_i^y$ to zero yields the dispersion relations for the spin wave.

We calculate the zero-temperature dynamic structure factor using Holstein-Primakoff
bosons. 

\begin{equation}
S(\mathbf{k}, \omega)=\sum_f \sum_{i=x,y,z} |<f|S^i (\mathbf{k})|0>|^2 \delta (\omega-\omega_f)
\end{equation}
Here $|0>$ is the magnon vacuum state and $|f>$ denotes the final state of
the spin system with excitation energy $\omega_f$.
Since $S^z$ does not change the number of magnons, it leads to the elastic part of the 
structure factor. 
Single magnon excitations contribute to the inelastic response through 
$\mathbf{S}^{x}(\mathbf{k})$ and $\mathbf{S}^{y}(\mathbf{k})$.

\section{Results for Vertical Stripes \label{results}}

\begin{figure}
{\centering \subfigure[$VS4$]{\resizebox*{!}{0.33\columnwidth}{\includegraphics{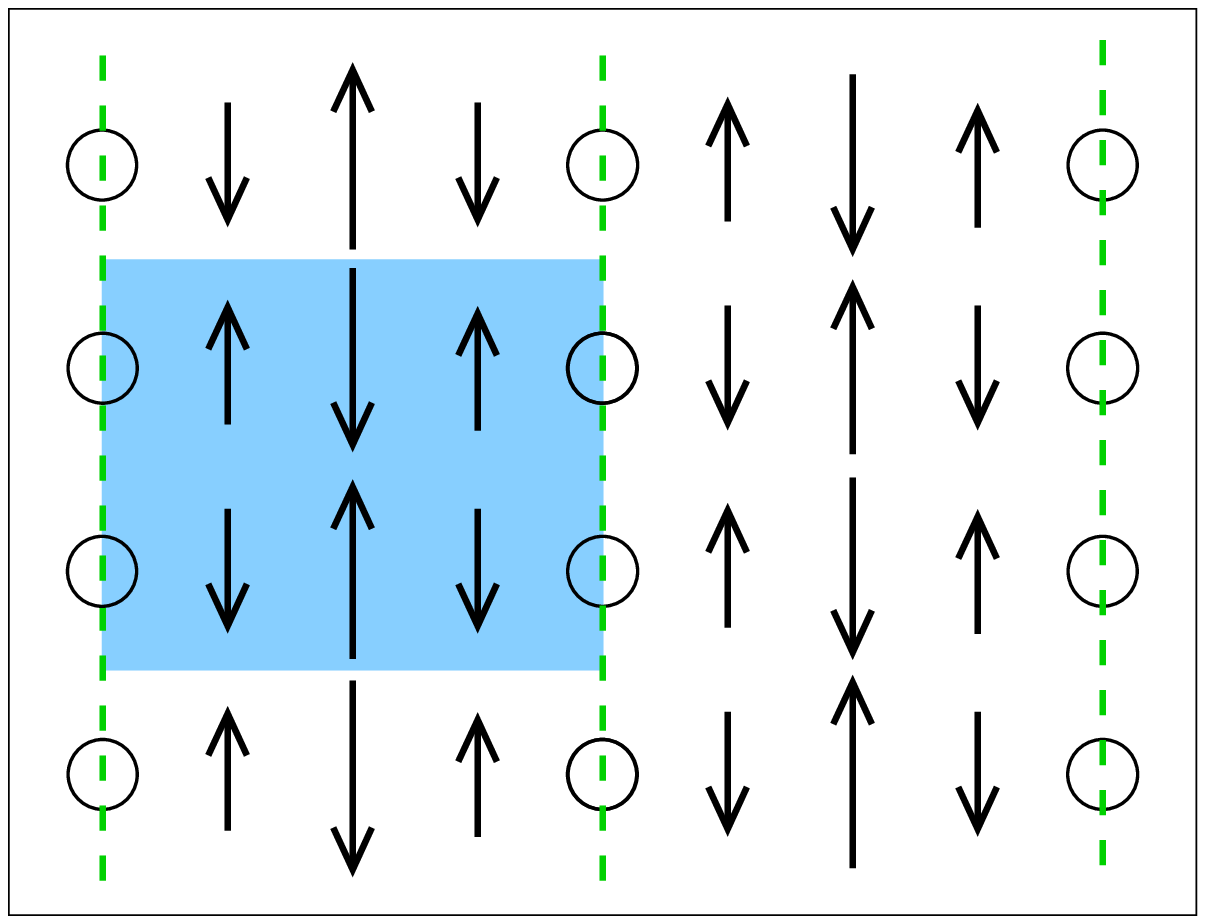}}} 
\subfigure[$VS5$]{\resizebox*{!}{0.33\columnwidth}{\includegraphics{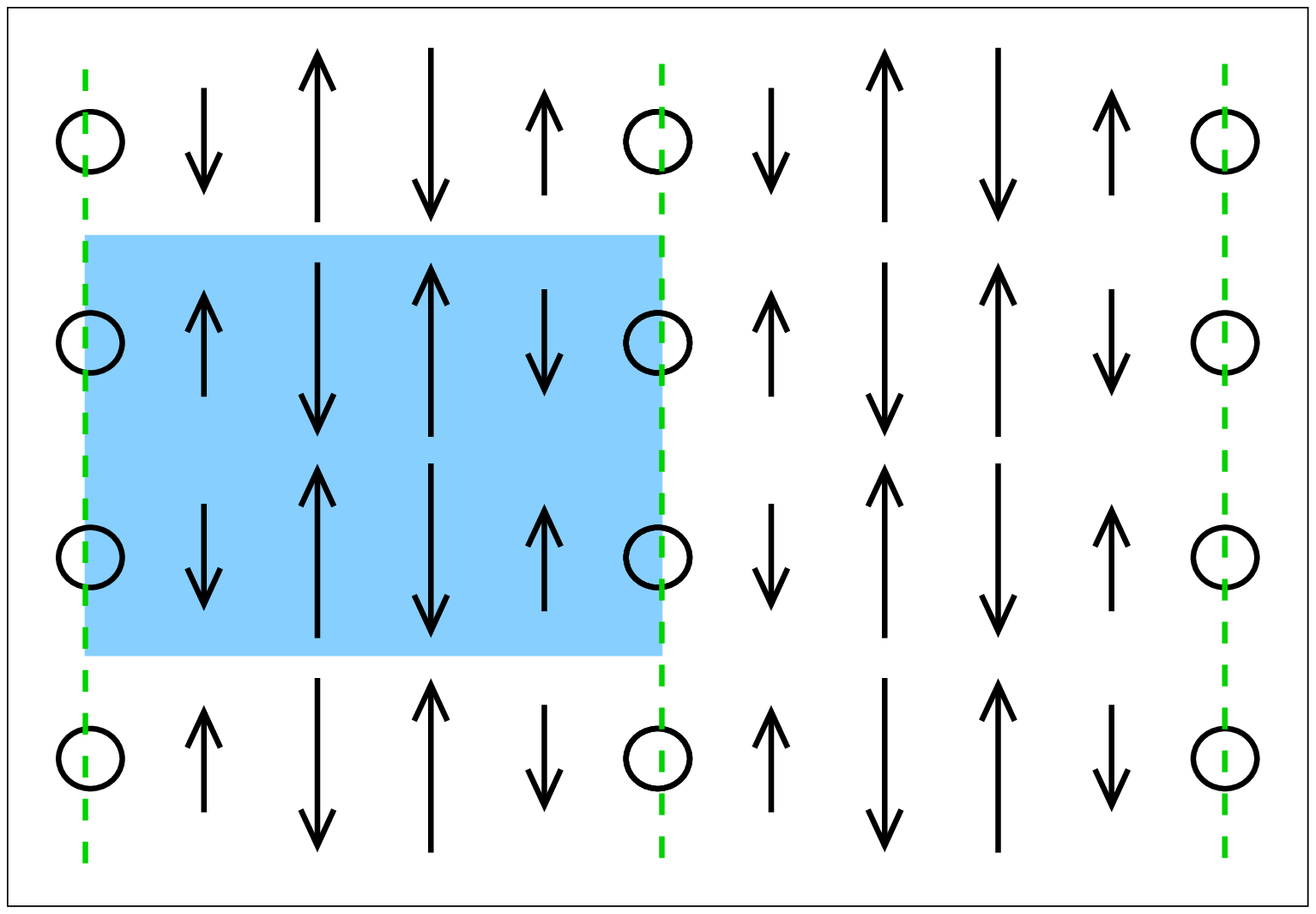}}} \par}

{\centering \subfigure[$VB4$]
{\resizebox*{!}{0.33\columnwidth}{\includegraphics{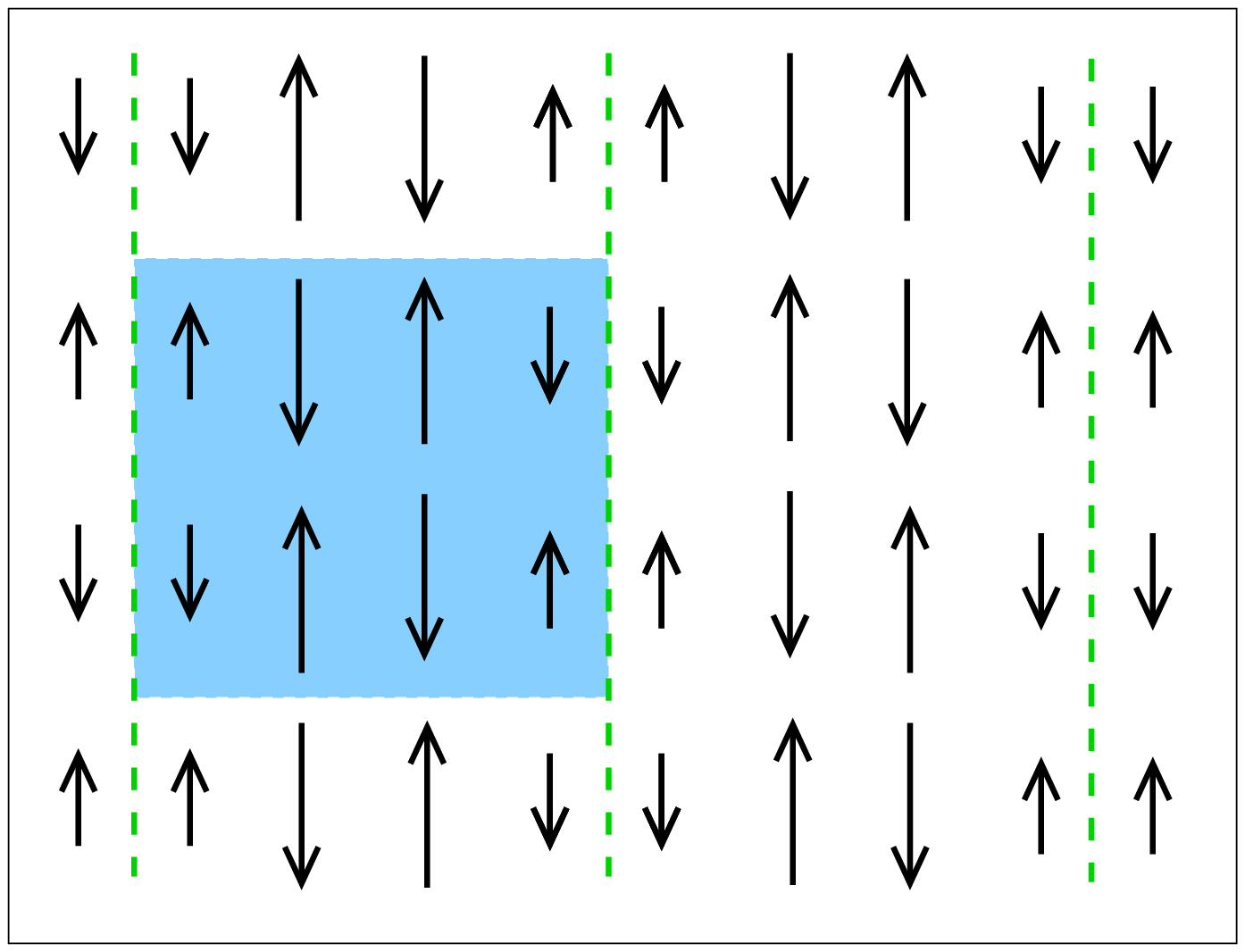}}} 
\subfigure[$VB5$]{\resizebox*{!}{0.33\columnwidth}{\includegraphics{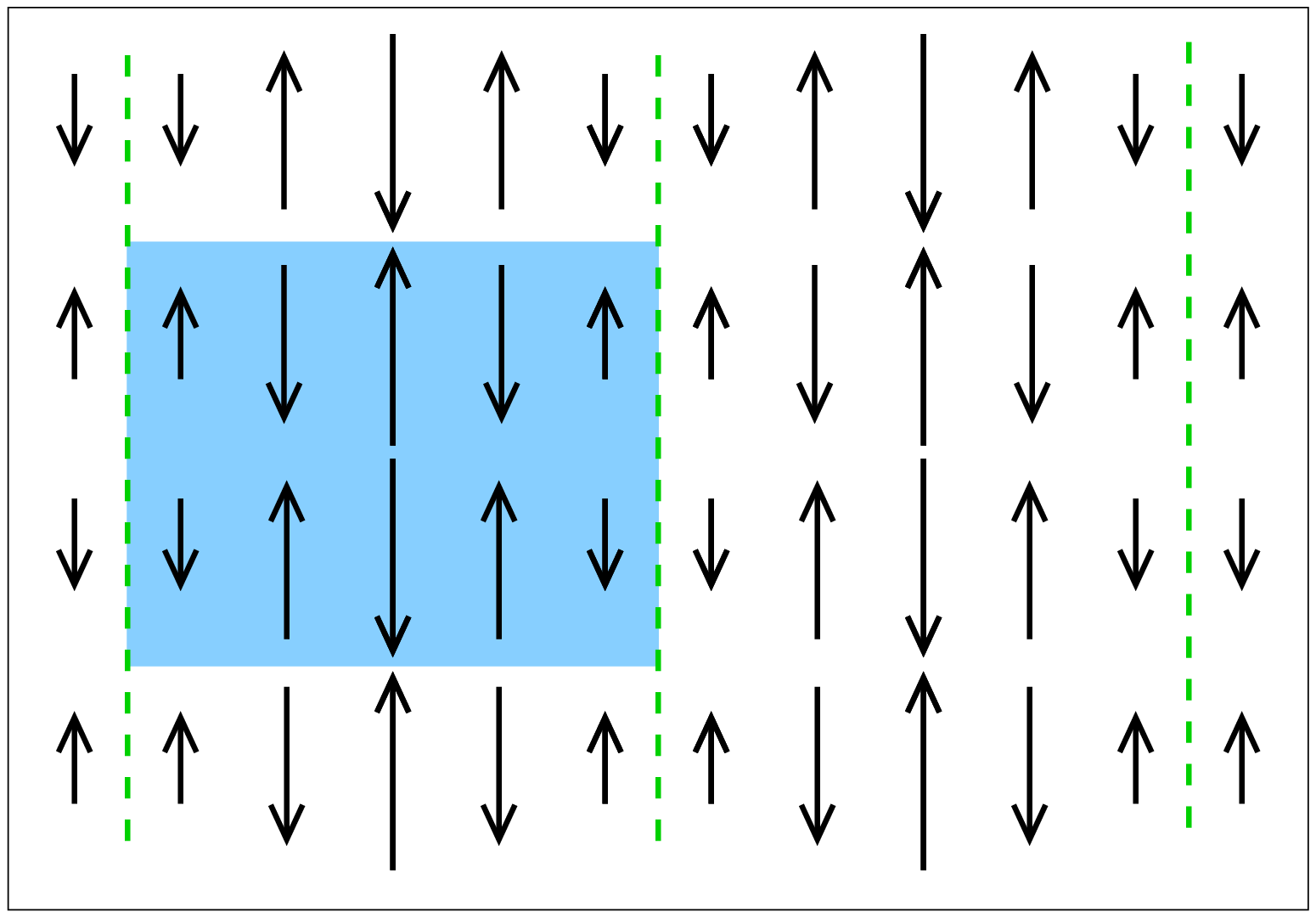}}} \par}

\caption{(Color online) Vertical site- and bond-centered configurations, showing even and odd spacing. 
``$S$'' refers to site-centered configurations, and ``$B$'' refers to bond-centered configurations.
The number label is the spacing $p$ between domain walls.
Dotted vertical lines mark antiphase domain walls.  The solid boxes denote unit cells.
The height of the arrows represents the net spin on a site, which is expected to peak between domain walls.
\label{vertical}}
\end{figure}

We begin with our results for ordered, vertical stripe phases.  We discuss magnon 
excitation energies
as functions of momentum, the dynamic spin structure factors, the elastic response,
the velocities of the acoustic bands, and analytic results for dispersion relations
for small unit cell sizes.
Fig.~\ref{vertical} shows schematic representations of vertical stripes that are
site- and bond-centered, with both even and odd spacing.  
In this figure (in contrast to Fig.~(\ref{couplings}))
we have used the length of the arrow to represent the net spin on 
a site.  The net spin is expected to be smaller near domain walls 
(as it is always zero on a domain wall).  Our zero frequency results
incorporate this general form factor. For the finite $\omega$ results,
we use a form factor
with the same net spin on each occupied site.

\subsection{Elastic peak at $(0,\pi)$ \label{elastic}}
\begin{figure}[hbt]
\begin{tabular}{cc}      
\subfigure[$VS5$]{\label{fig:VS5.form}\epsfig{file=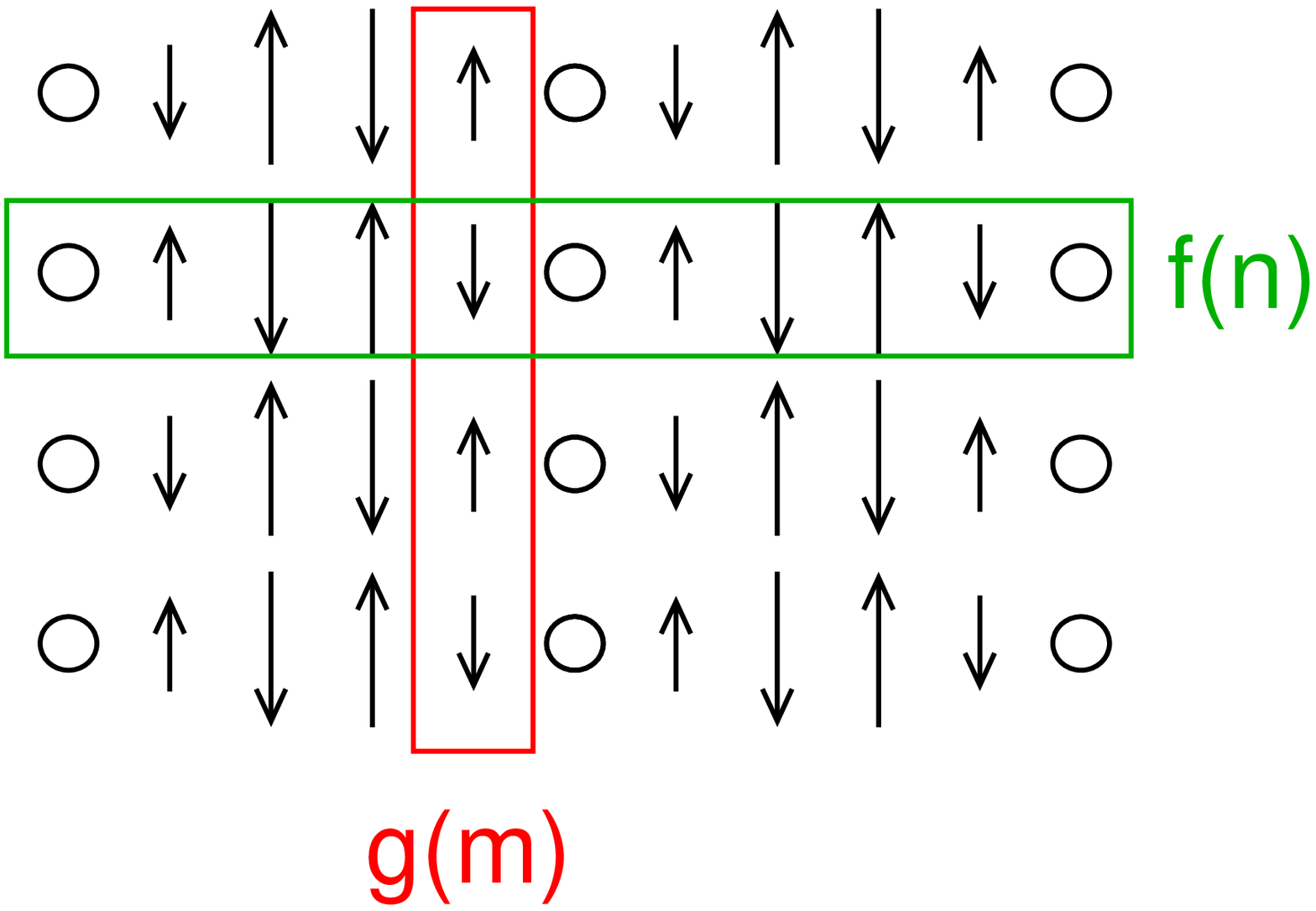, height=
0.43\columnwidth}} \\
\subfigure[$VB5$]{\label{fig:VB5.form}\epsfig{file=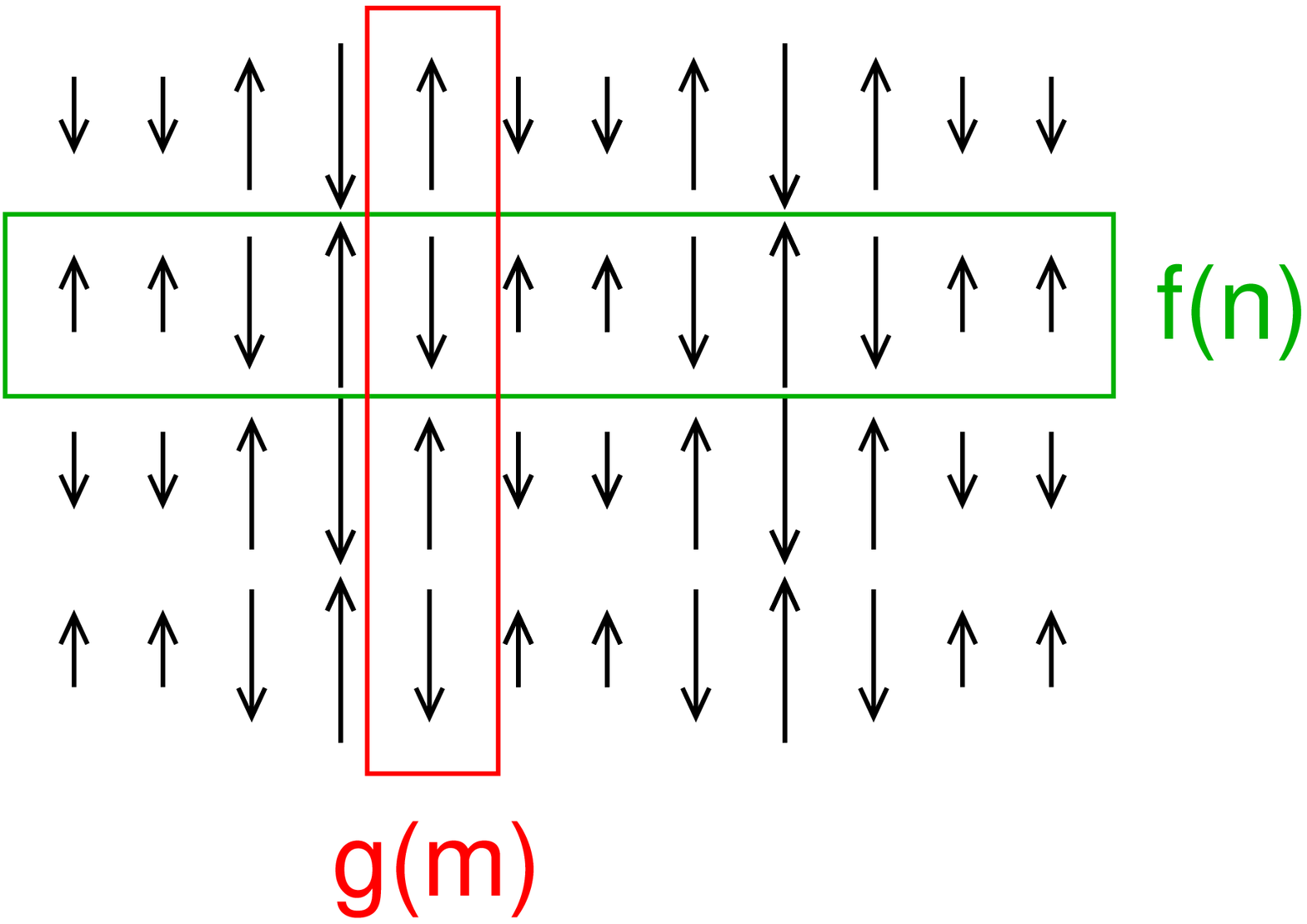, height=0.43\columnwidth}} \\ 
\end{tabular}
\caption{(Color online) Schematic representation of vertical stripes with $p=$odd widths,
indicating the pattern of the functions $g(m)$ and $f(n)$.
Note that for the bond-centered case with odd stripe spacings,
the function $f(n)$ can have a net magnetization, producing elastic weight
at the peak $(0,\pi)$.  
\label{fig:fg}}
\end{figure}

Elastic neutron scattering can in principle detect one important 
qualitative difference between bond- and site-centered
stripes. For odd stripe spacings, both bond- and site-centered stripes 
have magnetic reciprocal lattice vectors at $(0,\pi)$.
However, site-centered stripes are forbidden from producing weight
at $(0,\pi)$, whereas bond-centered stripes generically show weight at this point.
This is related to the discrete Fourier transform of the spin structure.
Taking advantage of the antiferromagnetic long range order in 
one direction and the finite spacing between stripes in the other, 
we can describe the spin structure in real space by a function
\begin{eqnarray}
S^{z}(n,m)&=&cos(\pi m)\sum_{j=0}^{j^{\prime}}A_{j}e^{i{2 \pi \over p}j n} \nonumber \\
&=&f(n)g(m)~,  \\
\label{fourier}
\end{eqnarray}
where $m$ is the discrete $y$ coordinate parallel to the stripes, 
$n$ is the discrete $x$ coordinate perpendicular to them,
and where $j^{\prime}=p-1$ for $p$ odd, with $j^{\prime}=2p-1$ for $p$ even.
The functions $f(n)$ and $g(m)$ are shown schematically in Fig.~\ref{fig:fg}.
The elastic scattering cross section is proportional to the
Fourier transform of $S^{z}(n,m)$:\cite{squires}
\begin{eqnarray}
\bigg(\frac{d\sigma}{d\Omega} \bigg)_{el}&\propto&\sum_{m,n} 
e^{i(k_m m + k_n n)}\big<S^{z}(m,n)\big> \big<S^{z}(0,0)\big>   \\
&=&\sum_m e^{i k_m m}cos(\pi m) \sum_{j=0}^{j^{\prime}}A_j \sum_n e^{i k_n n} e^{i
  ({2\pi j \over p})n}  \nonumber \\
&=&N_m (\delta_{k_{m},\pi}+\delta_{k_{m},-\pi}) 
\sum_{j=0}^{j^{\prime}}  
A_j \sum_n N_n \delta_{k_n,-{2\pi j\over p}}~.  \nonumber
\end{eqnarray}
We emphasize that this expression allows for 
{\em any form factor} and is not restricted
to configurations where each occupied site has a full quantum of spin.
In the case where each occupied site has the same net spin,
the ratio of intensity at the main peaks $(\pi \pm \pi/p,\pi)$ to that
at $(0,\pi)$ is $2$ in the VB3 case, and $2.6$ in the VB5 case.
Site-centered stripes always have $A_{0}=0$, while $A_{0}$ is generically
nonzero for bond-centered stripes (although it can be fine-tuned to zero).  
A finite $j=0$ term produces elastic weight at $(0,\pi)$.
This can be understood heuristically from considering the function
$f(n)$, shown schematically for the VB5 case in Fig.~\ref{fig:fg}.
Odd-spacing bond-centered stripes
generically have a net magnetization in the function $f(n)$,
while
symmetry forbids this for site-centered stripes.  

\subsection{Analytic Results for small $p$ \label{v.analy}}

For small $p$, which corresponds to small unit cell sizes, we can obtain
analtic results for the dispersion relation of the acoustic mode. 
For the case VS3, we find
\begin{equation}
\bigg({\omega_{VS3} \over {J_a S}} \bigg)^2 = 4(\lambda+1) + C - 4(\lambda+1) D
\label{VS3omega}
\end{equation}
where
\begin{eqnarray}
\lambda&=&|\frac{J_b}{J_a}| \\ \nonumber
C&=&2\lambda f(3 k_x) + 8 f(k_y) - 4 f^{2}(k_y) \\ \nonumber
D&=&|1-f(k_y)| \sqrt{1 - {2 \lambda f(3 k_x) \over (\lambda+1)^2}}
\end{eqnarray}
and the function $f$ is defined as
\begin{equation}
f(x) = 1-\cos(x)~.
\end{equation}

The acoustic spin wave velocity parallel to the stripe direction $(\mathbf{k}||\hat{y})$
may be obtained by setting $k_x =0$ above, and taking $k_y << 1$.  In this case, 
$f(k_x)=0$, $f(k_y) \rightarrow \frac{1}{2} k_y^2$, and 
$\omega_{VB2} \rightarrow v_{\parallel}|k_y|$, where
\begin{equation}
v_{\parallel}= \frac{1}{2} \sqrt{\lambda+3} ~v_{AF}, 
\end{equation}
and $v_{AF}=2\sqrt{2} J_a S$ is the velocity of the pure antiferromagnet with
coupling $J_a$ and no antiphase domain walls.
The spin wave velocity perpendicular to the stripe direction may
be similarly obtained:
\begin{equation}
v_{\perp}= \frac{3 \sqrt{2}}{4} \sqrt{\frac{\lambda(\lambda+3)}{\lambda+1}} ~v_{AF} .
\end{equation}
For $\lambda >>1$, these approach $v_{\perp}\rightarrow (3/\sqrt{2})\sqrt{\lambda}~v_{AF}$ and 
$v_{\parallel}\rightarrow
(1/2)\sqrt{\lambda}~v_{AF}$.  

For the case VB2, the problem reduces to diagonalizing a $4 \times 4$ matrix,
with the result 
\begin{equation}
\bigg( {\omega_{VB2} \over {J_a S}} \bigg)^2 = 2 (\lambda^2 + 3\lambda + 2) + A
- 2 \sqrt{(\lambda^2 + 3\lambda + 2)^2 + B}
\label{eqn:b2omega}
\end{equation}
where
\begin{eqnarray}
A&=&2 f(2 k_y) \\ 
B&=&-{1 \over 2}\lambda^2 f(4 k_x) - 4 f(k_y) -4(\lambda^2 + 3 \lambda) f(2 k_x)\nonumber \\
&&-4 f(k_y)(1 - f(k_y) + (\lambda^2 + 3\lambda)(1 - f(2 k_x))) \nonumber .
\end{eqnarray}

The spin wave velocities in the case VB2 are
\begin{equation} 
v_{\parallel} = {\sqrt{3} \over 2}~v_{AF}~,
\end{equation}
independent of $\lambda$, and 
\begin{equation}
v_{\perp} = \sqrt{ 3\lambda \over 2(\lambda +1)}~v_{AF}~.
\end{equation}
For $\lambda >> 1$, we note that $v_{\perp}$ {\em saturates} at
\begin{equation}
v_{\perp} \rightarrow \sqrt{3 \over 2}~v_{AF}~.
\end{equation}
That $v_{\perp}$ saturates with large $\lambda$ is in contrast to the behavior of 
site-centered cases and can lead to rather isotropic spin wave cones for the bond-centered 
case,\cite{john,boothroyd} despite local microscopic anisotropy.
As discussed in the next section, for bond-centered stripes with any spacing $p$, 
$v_{\parallel}$ is independent of 
$\lambda$ and $v_{\perp}$ saturates with large $\lambda$.

\subsection{Numerical Results}

\begin{figure}
\resizebox*{0.88\columnwidth}{!}{\includegraphics{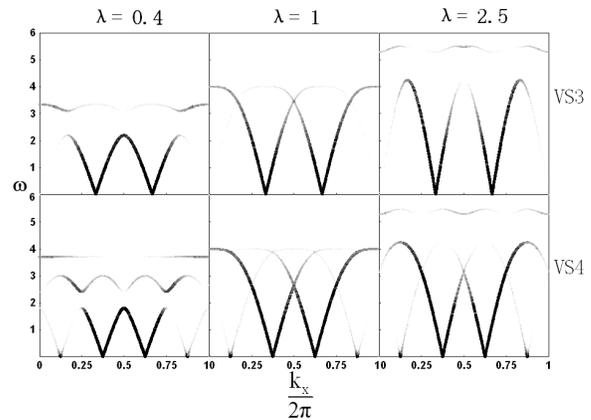}}
\caption{Spin wave spectra and intensities for vertical, site-centered stripes.
All spectra are reported at 
$k_y = \pi$ as a function of the transverse
momentum $k_x$. The frequency $\omega$ is in units of $J_a S$. Apparent crossings only occur at $\lambda = 1$ and $\lambda = 2.5$~.\label{spectra.vs}}
\end{figure}

\begin{figure}
\resizebox*{0.88\columnwidth}{!}{\includegraphics{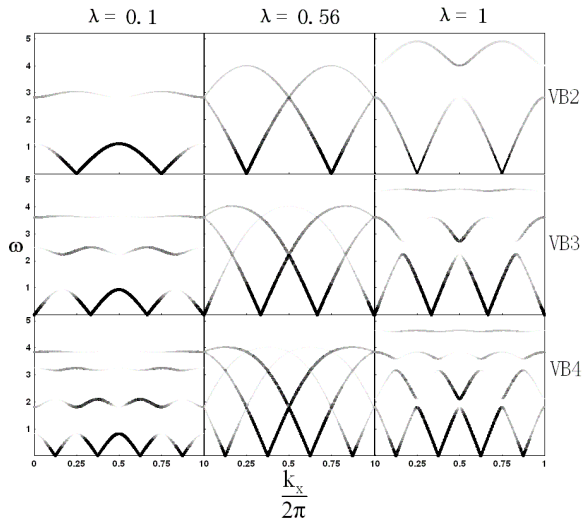}}
\caption{Spin wave spectra and intensities for vertical, bond-centered stripes.
All spectra are reported at 
$k_y = \pi$ as a function of the transverse
momentum $k_x$.  The frequency $\omega$ is in units of $J_a S$.\label{spectra.vb}}
\end{figure}

For most values of the stripe spacing $p$, the spin wave matrices are sufficiently large that
one must use numerical diagonalizations to obtain the 
dispersion relations of the various modes. From the corresponding eigenfunctions we can then also 
calculate the spectral intensity (proportional to the dynamic structure factor) that these magnon states
would contribute to the inelastic neutron
scattering.  
Figs.~\ref{spectra.vs} and \ref{spectra.vb} show the calculated dispersion and scattering intensities for site- 
and 
bond-centered vertical stripes  of various spacings. 
Our results for site-centered stripes are consistent with those 
of Ref.~\cite{kruger}.  
For the site-centered case, bands never cross for $\lambda < 1$.
At the critical couplings $\lambda = 1$ and $\lambda = 2.5$, site-centered bands appear to cross.
Away from these couplings, vertical site-centered bands generally repel rather than cross.
For $\lambda = 1$, the dispersion is very similar
to that of a pure antiferromagnet, albeit with different magnetic
reciprocal lattice vectors.  For any coupling $\lambda$, as 
$p \rightarrow \infty$, the result for a pure 2D antiferromagnet 
is recovered.  For $p$ increasing but finite, the number of
bands as well as the number of reciprocal lattice vectors increases.
However as $p \rightarrow \infty$, all spectral weight is transferred
to the response of a pure antiferromagnet.

Fig.~\ref{spectra.vb} shows representative results for vertical bond-centered
antiphase domain walls with spacings $p=2, 3,$ and $4$. In this case,
the critical point where bands appear to touch is at $\lambda_c \approx 0.56$
and is at most only weakly dependent on $p$.
Away from the critical coupling, bands never appear to cross,
but rather level repulsion is observed.  
There are other notable differences between the site- and bond-centered cases.
For one thing, for the same spacing $p$, bond-centered configurations 
yield one more band: site-centered configurations have $p-1$ bands, whereas
there are $p$ bands for bond-centered configurations.

A qualitative difference between the two cases is the scaling of the
band energies with coupling $\lambda$.  For site-centered configurations,
all bands increase their energy monotonically with the coupling ratio
$\lambda$. 
This is in contrast with the bond-centered case, where for large $J_b$, only the top band
is affected by the ferromagnetic coupling (that is, it increases linearly with $\lambda$), 
but all other bands saturate as $\lambda$ is increased.
The behavior of the top band can be understood by considering the 
spins that are ferromagnetically coupled across the domain wall.  
In the top band, these spins precess $\pi$ out of phase with each other,  
and the 
dispersion is dominated by the behavior of 
the effective ferromagnetic dimers, yielding 
$\omega \rightarrow 2 |J_b| S/\hbar$ as $|J_b| \rightarrow \infty$,
as shown in Appendix~\ref{dimer}.

An important consequence of the saturation of the lower bands as $\lambda$ gets large
in the bond-centered cases is that  
the low-energy spin wave velocities alone, $v_{\perp}$ and $v_{\parallel}$, cannot readily
be used to extract the relation between the 
bare exchange couplings $J_a$ and $J_b$.  We explore this point in more detail
in the next section.

\begin{figure}
{\centering \subfigure[Acoustic spin wave velocities for VS3]
{\resizebox*{\columnwidth}{!}{\includegraphics{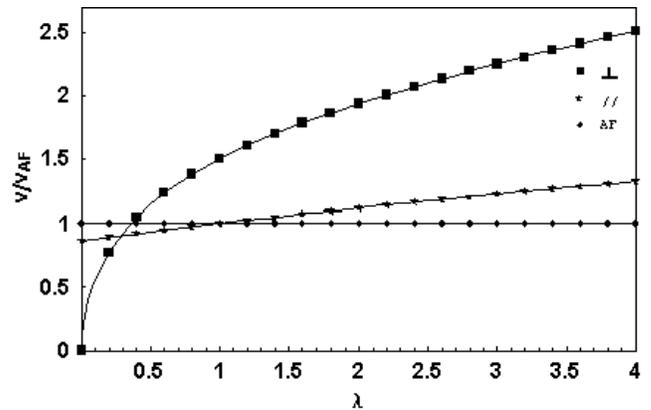}}} \par}

{\centering \subfigure[Acoustic spin wave velocities for VB3]{\resizebox*{\columnwidth}{!}{\includegraphics{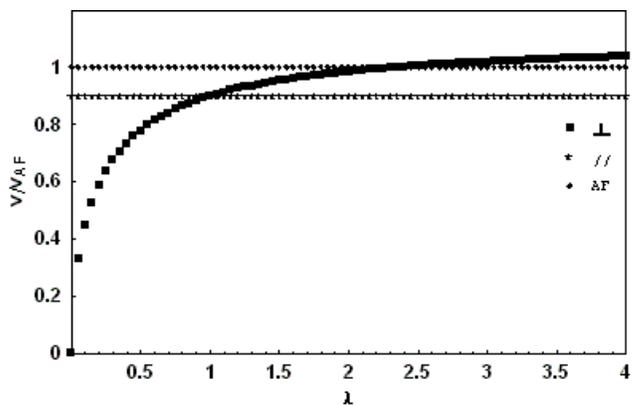}}} \par}

\caption{Spin wave velocities for (a) VS3 and (b) VB3 as a function of 
the coupling ratio $\lambda$. 
The solid curves in panel (a) are analytic results for VS3 calculated in
Section \ref{v.analy}.  
Symbols in both (a) and (b) are numerical results.  
The velocities parallel and perpendicular to the stripe direction are
equal to each other for $\lambda = 2/7$ and $\lambda = 1$ in the
site- and bond-centered cases respectively.  
Qualitatively similar behavior is found for other stripe spacings.
\label{velocity}}
\end{figure}

In Fig.~\ref{velocity} we present the spin wave velocities
perpendicular ($v_{\perp}$) and
parallel ($v_{\parallel}$) to the stripe orientation 
for the acoustic (lowest) bands
as functions of the coupling constant ratio $\lambda$.
These are compared to the reference velocity, $v_{AF}$ 
of the pure antiferromagnet,
which is independent of the coupling $\lambda $
and equivalent to $p\rightarrow \infty$. 

While in both the site- and bond-centered cases (Figs.~\ref{velocity}(a) and
(b), respectively) the perpendicular
velocity depends on the coupling ratio, in the bond-centered case
$v_{\perp}$ rapidly saturates to a value close to $v_{\parallel}$ 
for large $\lambda$. As a consequence, the value of the 
coupling ratio $\lambda = J_b / J_a$ 
cannot be determined solely by the ratio of the acoustic velocities but requires
additional information, such as $v_{AF}$.

The curves of $v_{\perp}$ and $v_{\parallel}$ cross at $\lambda = 1$ 
for the bond-centered case, apparently independent of $p$ for the widths we have studied.  
The crossing is at most weakly dependent on $p$ in the site-centered case,
occurring at $\lambda = 2/7$ for DS3, and at $\lambda = 0.3$ for DS4.  
For all spacings studied, we find that in the bond-centered case, $v_{\parallel}$
is independent of the coupling $\lambda$ 
and that $v_{\perp}$ rapidly saturates with large $\lambda$.
As $p$ gets larger, both of these 
velocities approach $v_{AF}$.
For the VB3 configuration, $v_{\parallel} = 0.9 v_{AF}$, independent of $\lambda$.
Notice that the independence of $v_{\parallel}$ upon
$\lambda$ and the rapid saturation of $v_{\perp}$ as $\lambda$ 
becomes larger than $1$ means that bond-centered configurations
can produce rather isotropic spin wave cones.\cite{john,boothroyd}

\section{Results for Diagonal Stripes}

\begin{figure}
{\centering \subfigure[$DS4$]{\resizebox*{!}{0.3\columnwidth}{\includegraphics{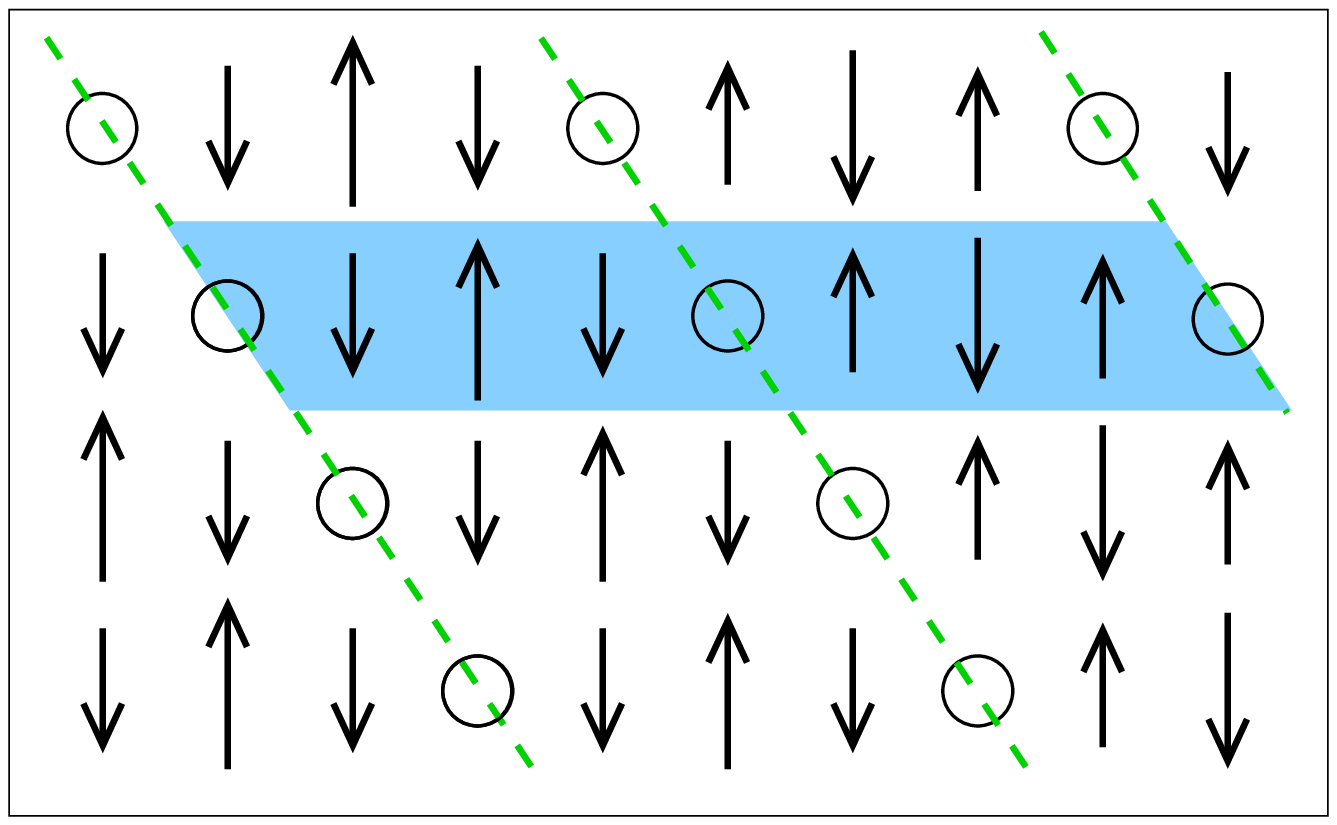}}} 
\subfigure[$DS5$]{\resizebox*{!}{0.3\columnwidth}{\includegraphics{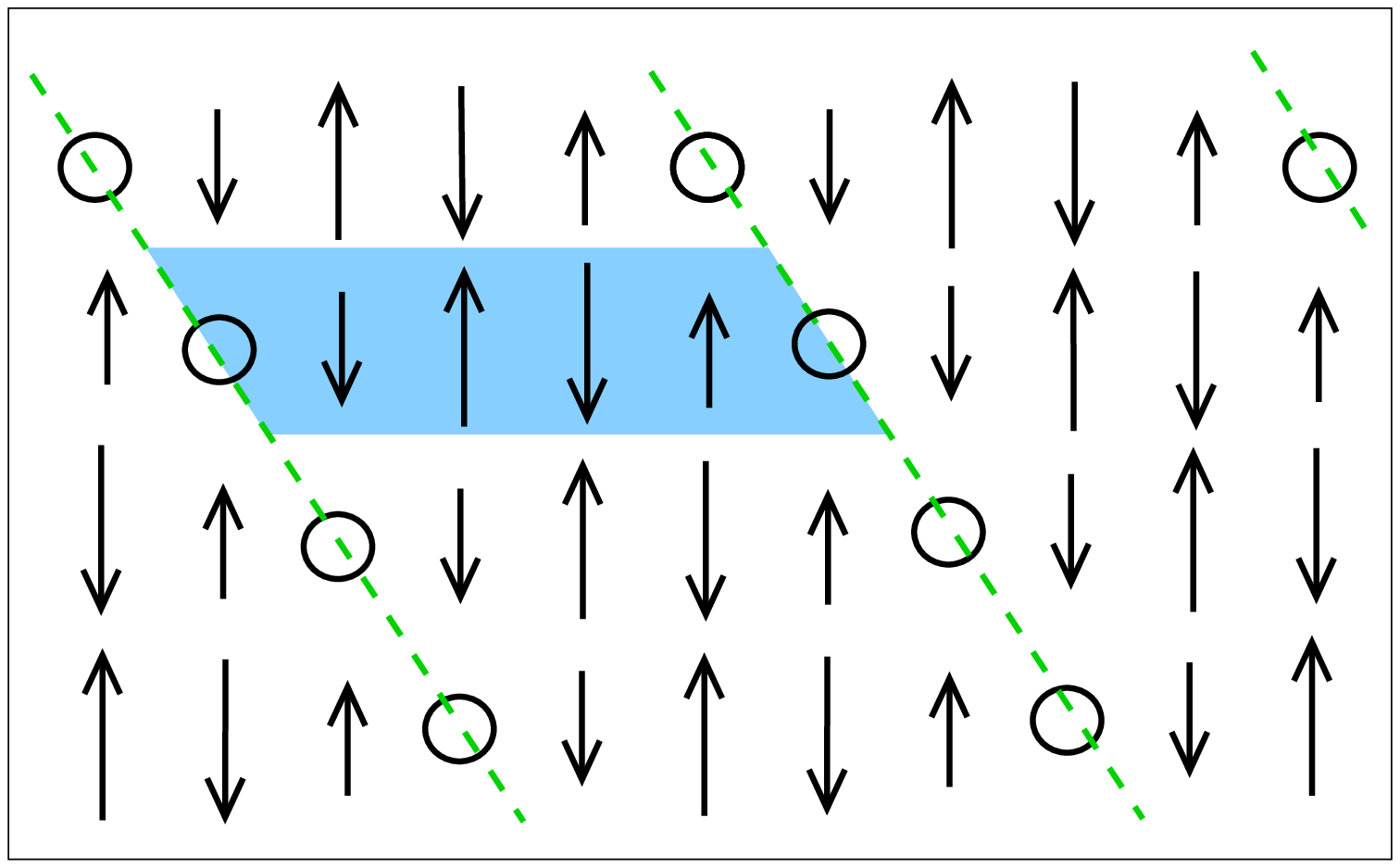}}} \par}

{\centering 
\subfigure[$DB4$]
{\resizebox*{.45\columnwidth}{0.31\columnwidth}{\includegraphics{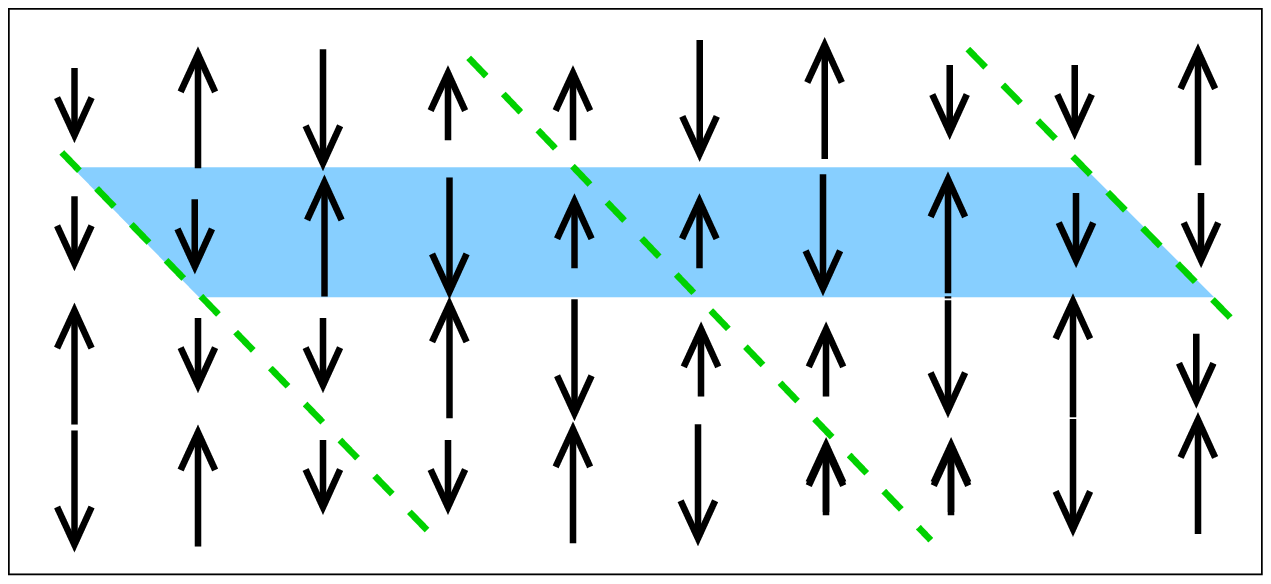}}} 
\subfigure[$DB5$]{\resizebox*{.45\columnwidth}{0.3\columnwidth}{\includegraphics{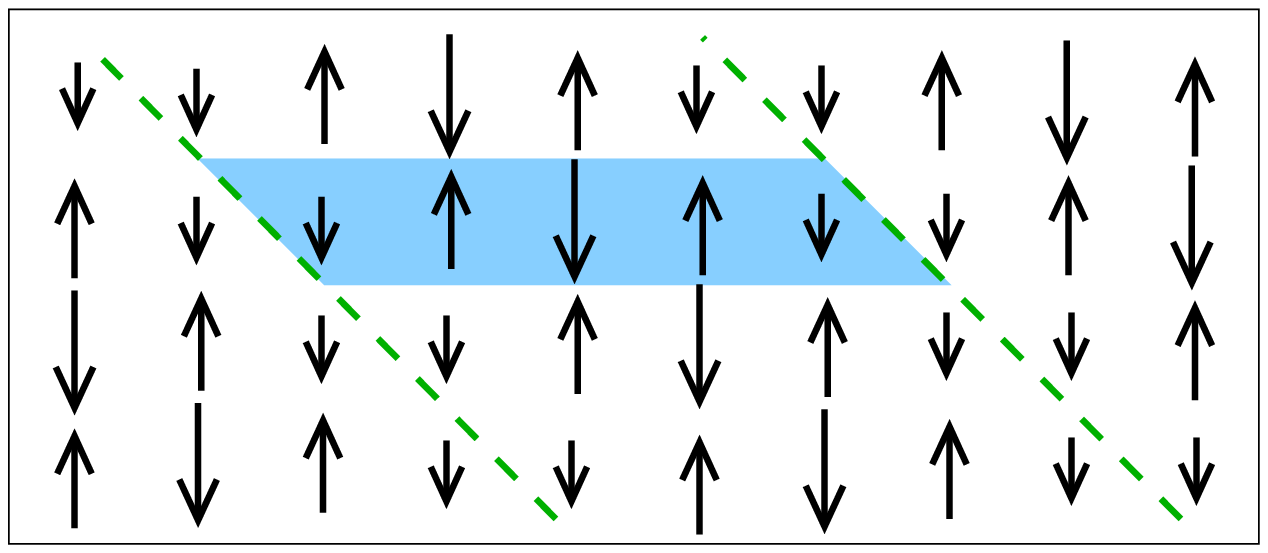}}} \par}

\caption{(Color online) Diagonal site- and bond-centered configurations, showing even and odd spacing. 
Dotted lines denote domain walls.  Solid parallelograms denote unit cells.\label{diagonal}}
\end{figure}

Fig.~\ref{diagonal} depicts representative
diagonal configurations, for site- and bond-centered domain walls and with even and odd
spacing.  As mentioned in Sec~\ref{sec:diag}, for a given microscopic model, 
diagonal stripes are more strongly
coupled across the domain wall than vertical
stripes.  In addition,
there are more parameters to consider for site-centered diagonal stripes: we
must include  $J_c$ as well as  $J_b$ (see Fig.~\ref{couplings}), 
since both couplings appear to the same order
if derived from, {\em e.g.}, a Hubbard-like model.  

\subsection{Elastic Peak at $(0,0)$}
Like their vertical counterparts, 
{\it bond-centered} diagonal stripes can produce new peaks in the elastic response.
With diagonal stripes the new weight is physically transparent.
For all bond-centered domain walls, nearest-neighbor spins are ferromagnetically coupled 
across the wall, 
and in the diagonal case, nearest neighbor pairs {\it along} a single domain wall all
have their moments pointing in the same direction,  
leading to a domain wall magnetization.
As Fig~(\ref{diagonal}) illustrates, for diagonal stripes with even $p$, 
adjacent domain walls have alternating signs of the 
magnetization.
But diagonal stripes with odd spacing have the same magnetization direction
on each domain wall.  This generically leads to net ferromagnetism 
and a peak at $(0,0)$, unless parameters are fine-tuned.
In a three dimensional antiferromagnet, (as may happen, \cite{sitecentered} {\em e.g.,} with
weakly coupled planes) domain walls are two dimensional (planar), 
and this peak appears either at $(0,0,\pi)$ if the diagonal in-plane
stripes lie directly on top of each other from plane to plane
(meaning there is also no net magnetization on a domain wall),
or at $(0,0,0)$ if the stripes are diagonal within a plane and in their 
correlation from plane to plane.

\subsection{Analytic results for small $p$}

As for the case of vertical stripes discussed in Sec.~\ref{v.analy}, for small $p$, it 
is possible to obtain analytic
forms for the acoustic dispersion relations for diagonal stripes in both the site- and
bond-centered cases.  

For the case DS3, the analytic dispersion is
\begin{eqnarray}
\big{(}\frac{\omega_{DS3}}{J_{a}S}\big{)}^2/2 &=& f(k_{x} - k_{y}) + \lambda^2\,f(2\,\left( k_{x} - k_{y} \right) )  \nonumber \\
&+&  2\,\lambda \lambda_{c}\,f(k_{x} - k_{y})   \nonumber \\
&+& (\lambda+\lambda_{c})\big[f(2\,k_{x} + k_{y}) +  f(k_{x} + 2\,k_{y})\big] \nonumber \\
&+& \lambda\,\big[ f(3\,k_{x}) +  f(3\,k_{y}) \big]~, \\
\end{eqnarray}
where $\lambda_c=|\frac{J_c}{J_a}|$, and
where the function $f$ is defined as
\begin{equation}
f(x) = 1-\cos(x)~,
\end{equation}
as in Sec.~\ref{v.analy}.

The dispersion perpendicular to the stripes, along the $\mathbf{k}=(k_x,k_x)$
direction, is then
\begin{equation}
\frac{\omega (k_x,k_x)}{J_a S}=  \sqrt{8(2 \lambda+\lambda_c)}~
\big{|}\sin{(\frac{3 k_x}{2})}\big{|},
\end{equation}
which yields for the velocity in that direction
\begin{equation}
v_{\perp}=\frac{3 \sqrt{2\lambda+\lambda_c}}{2 \sqrt{2}}~v_{AF},
\end{equation}
which approaches $v_{\perp} \rightarrow {3 \over 2}\sqrt{2 \lambda}~v_{AF}$
for large $\lambda$, and $v_{\perp} \rightarrow {3 \over 2}\sqrt{\lambda_c}~v_{AF}$
for large $\lambda_c$.

In the parallel direction $(k_x,-k_x)$, the dispersion becomes
\begin{eqnarray}
{1 \over 8}\bigg(\frac{\omega (k_x,-k_x)}{J_a S}\bigg)^2  &=&   (1+\lambda+\lambda_c+\cos{k_x}+\lambda\cos{(2k_x)}) \nonumber \\
& \times &  (1+2\lambda+2\lambda
  \cos{k_x})  \sin^2{\frac{k_x}{2}}~, \\
\end{eqnarray}
which gives
\begin{equation}
v_{\parallel}= \frac{\sqrt{(1+4\lambda)(2+2\lambda+\lambda_c)}}{2\sqrt{2}}~v_{AF}.
\end{equation}
This approaches $v_{\parallel} \rightarrow \lambda~v_{AF}$ for large $\lambda$,
and $v_{\parallel} \rightarrow \sqrt{\lambda_c/8}~v_{AF}$ for large $\lambda_c$.

For the case DB2, the analytic dispersion is

\begin{equation}
\big{(}\frac{\omega_{DB2}}{J_a S}\big{)}^2/2 = 4 \lambda (1 + \lambda) + A - \lambda \sqrt{16(1 + \lambda)^2 + B}
\end{equation}
where
\begin{eqnarray}
A&=&(1-\lambda^2) f(k_{x}-k_{y})  \\
B&=&-8(1+\lambda)^2 f(k_{x}-k_{y}) \nonumber \\
&+&\big(2-f(k_{x}-k_{y})\big) \big[ 2f(k_{x}-k_{y}) \nonumber \\
&-& 2 f(2k_{x}-2k_{y}) -f(3k_{x}+k_{y}) - f(k_{x}+3k_{y}) \big]~. \nonumber
\end{eqnarray}
Perpendicular to the stripes, along $\mathbf{k} = (k_x,k_x)$, 
the velocity is
\begin{equation}
v_{\perp} = \sqrt{\lambda \over \lambda + 1}~v_{AF}~,
\end{equation}
saturating to $v_{\perp} \rightarrow v_{AF}$ as $\lambda >> 1$.
In the direction $\mathbf{k} = (k_x,-k_x)$, parallel to the stripes,
the velocity is 
\begin{equation}
v_{\parallel} = {1 \over 2}\sqrt{\lambda + 1}~v_{AF}~.
\end{equation}

\subsection{Numerical Results}

\begin{figure}
{\centering \resizebox*{1\columnwidth}{!}{\includegraphics{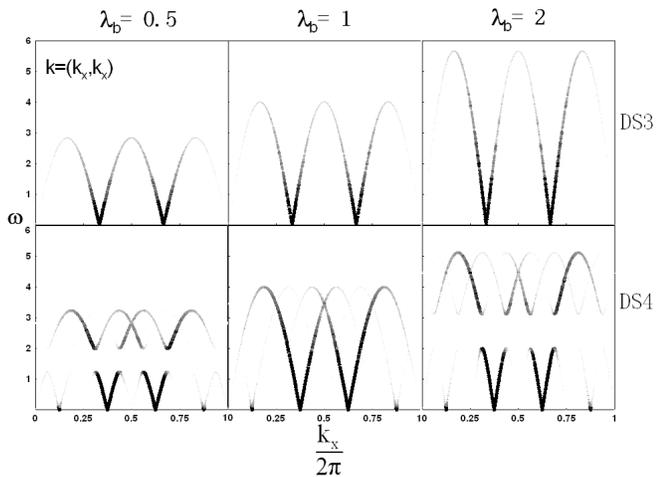}} \par}
\caption{Dispersion and intensities for DS3 and DS4 along ($k_x$,$k_x$) direction
  with $J_b$ only.   The frequency $\omega$ is in units of $J_a S$. For all plots, $J_{c} = 0$~. 
\label{specds34}}
\end{figure}

In Fig.~\ref{specds34}, 
we plot the dispersion and intensities for DS3
and DS4 along ($k_x$,$k_x$) for various values of the
coupling ratio $\lambda = |\frac{J_b}{J_a}|$,
setting $J_{c}=0$. (See Fig.~\ref{couplings} for the definitions of $J_b$ and $J_c$.)
Similar results using a $J_c$ only model ({\em i.e.} with $J_b = 0$)
are reported in 
Ref.~\cite{kruger}.
Our results show similar band structures but with critical
coupling $\lambda=1$, which is only half of the $J_c$ only model.  

\begin{figure}
{\centering \resizebox*{1\columnwidth}{!}{\includegraphics{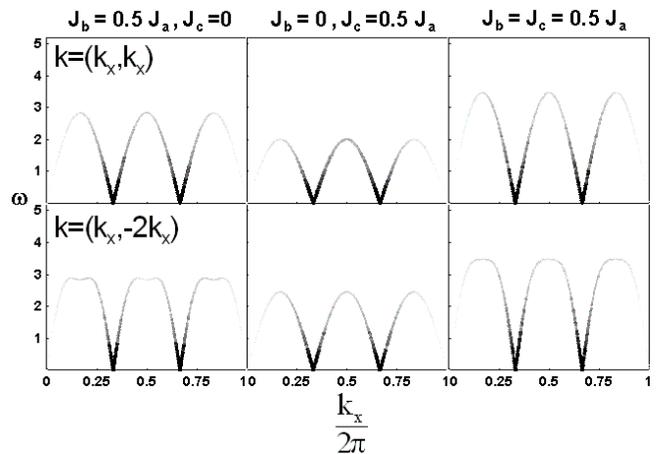}} \par}
\caption{Dispersion and intensities for DS3 along ($k_x$,$k_x$) and ($k_x$,$-2 k_x$)
  directions,  comparing the effects of $J_b$ and $J_c$.  The frequency $\omega$ is in units of $J_a S$.\label{compare}}
\end{figure}

However, the effects of $J_b$ and $J_c$ depend upon the direction in $k$ space.
In Fig.~\ref{compare} we  show the effects of varying the couplings $J_b$ and $J_c$
for two cuts in momentum space for DS3.  For a cut perpendicular to the stripe direction,
$J_b$ and $J_c$ have more or less the same effect, although since $J_b$ couples
more spins than $J_c$, it has a more dramatic effect.  Increasing either coupling 
broadens the bandwidth in a roughly linear manner with negligible effect on the shape.
However, for the cut $(k_x,- 2 k_x)$,
we see that the presence of $J_b$ produces inflection points when $J_{c}=0$,
and can produce flat-topped dispersions if $J_c$ is included as well.

\begin{figure}
{\centering \resizebox*{!}{1\columnwidth}{\includegraphics{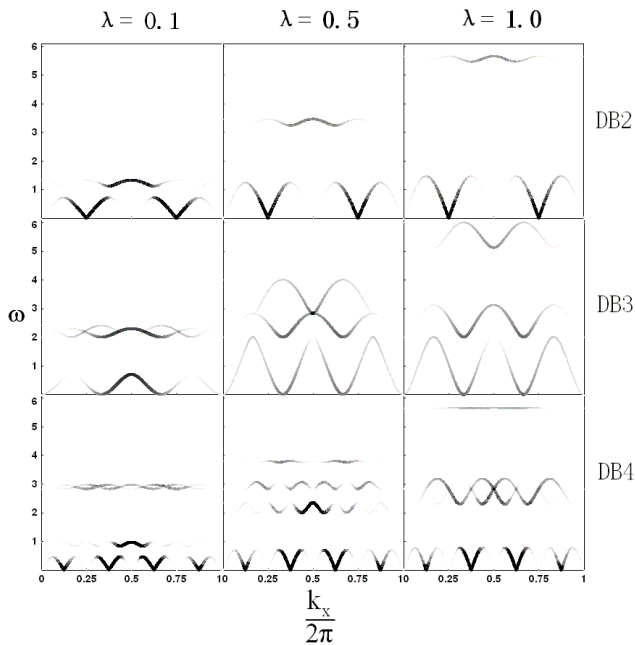}} \par}
\caption{Dispersion and intensities for diagonal bond-centered domain walls
  along ($k_x$,$k_x$) direction,
at $\lambda  = 0.1, 0.5, 1.0$, for $p=2, 3,$ and $4$~.  The frequency $\omega$ is in units of $J_a S$.
\label{dbond}}
\end{figure}

We show in Fig.~\ref{dbond} the calculated dispersion relations and intensities 
for the bond-centered diagonal case, for spacings $p=2,3,$ and $4$.  
As in the vertical case, the number of bands is equal to $p$.
A striking difference
in the spectra of odd spacings is seen, as the net ferromagnetism in the system
changes the low-energy character of the spin waves from a linear (antiferromagnetic-like)
to a quadratic (ferromagnetic-like) dispersion.  
Rather than the band repulsion observed in the vertical case (except at 
finely tuned values of the coupling), crossing of optical bands is generic in the
bond-centered diagonal case.
Note the ability of optical bands to cross, indicating a difference in symmetry
for the crossing bands.  Also evident in the dispersion of $DB4$ is the 
downturn of the acoustic band at $2p$ magnetic reciprocal lattice vectors, 
twice as many as in the odd case.
(See Sec.~\ref{sec:diag}~.)
This is expected because of the doubling of the unit cell necessary to
accommodate even spacing.
Note, however, that spectral weight is forbidden at these extra reciprocal lattice vectors, 
including the $(\pi,\pi)$ point.  

\begin{figure}
{\centering \resizebox*{\columnwidth}{!}{\includegraphics{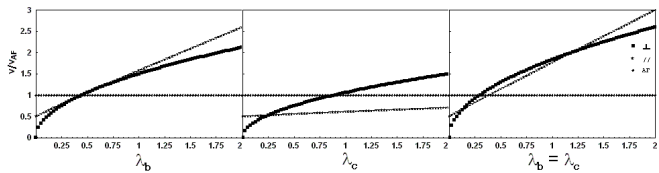}} \par}
\caption{Velocities parallel ($v_{\parallel}$) and perpendendicular ($v_{\perp}$)
to the stripe direction, as compared to $v_{AF}$, for DS3. 
In the first panel, $J_{c} = 0$, and the velocities are plotted as a function
of $\lambda_{b} = J_{b}/J_{a}$.  In the second panel, $J_{b}=0$, and 
the velocities are plotted as a function of $\lambda_{c} = J_{c}/J_{a}$.  
In the third panel, $J_{b}=J_{c}$, and the velocities are plotted
as a function of $\lambda_{b}=\lambda_{c}$.
\label{fig:vel.ds}}
\end{figure}

\begin{figure}
{\centering \resizebox*{1\columnwidth}{!}{\includegraphics{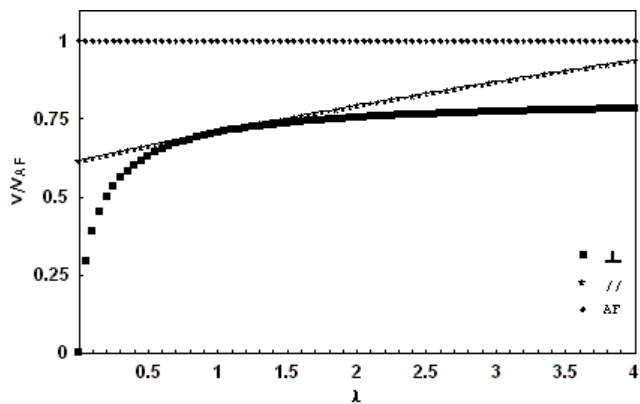}} \par}
\caption{Velocities parallel ($v_{\parallel}$) and perpendendicular ($v_{\perp}$)
to the stripe direction, as compared to $v_{AF}$, for DB4. \label{fig:v.db4}}
\end{figure}

In Fig.~\ref{fig:vel.ds}, we plot the spin wave velocities for $DS3$.
When $J_b$ and $J_c$ are both finite,
there is a wide range of couplings $\lambda$ for which the spin wave velocities
parallel and perpendicular to the stripes are nearly equal,
while this approximate isotropy is confined to a narrow range of $\lambda$ if either $J_b$ 
or $J_c$ is zero.    
Fig.~\ref{fig:v.db4}, which presents $v_{\perp}$ and $v_{\parallel}$ for the case DB4, shows the 
characteristic saturation of $v_{\perp}$ with large 
$\lambda$ for bond-centered stripes.

\section{Experimental Implications}

We have shown that for a certain class of nontrivial spin orderings on a lattice,
the spin wave response is sensitive to the microscopic placement of the antiphase domain walls.
Furthermore, even {elastic} neutron scattering can in principle distinguish site- 
from bond-centered for odd stripe spacings, whether vertical or diagonal.  

While both site- and bond-centered odd width
vertical stripe configurations
will produce elastic weight at $(\pi \pm {\pi \over p},\pi)$, 
only configurations that are phase-shifted
from the site-centered configuration ({\em e.g.} a bond-centered configuration) 
are capable of producing weight at $(0,\pi)$, and the 
observation of this peak along with peaks at 
$(\pi \pm {\pi \over p},\pi)$ would rule out a site-centered vertical configuration.  
A similar ferromagnetic peak--{\em i.e.}, at (0,0)--would rule out site-centered diagonal 
stripes.\cite{sitecentered}

Figs.~\ref{velocity} 
and~\ref{fig:v.db4} illustrate another important implication for experiments: The 
transverse spin wave velocity $v_{\perp}$ in the acoustic band {\em saturates} for 
large $\lambda$ in the
bond-centered case for both vertical and diagonal stripes.  
In fact, all but the top band in the bond-centered case saturate and become 
{\em independent} of 
$J_b$ for large $J_b$.  As noted above, this unfortunately means that an 
estimate of $\lambda = J_{b}/J_{a}$
cannot necessarily be discerned directly from the ratio 
$v_{\perp}/v_{\parallel}$ but
requires either independent knowledge of whether the stripes are 
site- or bond-centered, or
an appropriate estimate of the bare coupling (from, {\em e.g.} $v_{AF}$).  

A prominent piece of phenomenology in the cuprates is the 
the ``resonance peak'' observed in neutron scattering,\cite{keimerprl,mook2,resonance} 
which is the presence of extra scattering weight appearing at $(\pi,\pi)$ at 
finite frequency, typically of order $40meV$.  One proposal is that this may
be due to spin waves crossing.\cite{balatsky,balatsky01a} 
We note that for vertical stripes, spin waves generically repel 
and appear to cross only at finely tuned values of the
coupling.
For site-centered configurations, this corresponds to $\lambda = 1$ and $\lambda \approx 2.5$,
while for bond-centered configurations, the critical coupling is near $\lambda \approx 0.56$~.
However, a finite energy resolution measurement would not be able to distinguish 
actual crossings from near crossings.
In the bond-centered case with large $\lambda$, 
the first optical mode has more weight at $(\pi,\pi)$ than the acoustic band, 
which would tend to leave the weight near the elastic incommensurate peaks
$(\pi \pm {\pi \over p},\pi)$
rather disconnected from what might be called a ``resonance peak'' in this configuration.  
We also note that our calculations show that band crossings are more generic in the presence of diagonal stripes 
than vertical stripes.

The nickelate compound 
La$_{1.69}$Sr$_{0.31}$NiO$_4$ shows evidence from neutron scattering of diagonal stripes with spacing $p=3$.
\cite{john,boothroyd}
 As Sr  is substituted for La, holes are doped into the NiO$_2$ planes.  
Neutron scattering has been used to map out the acoustic spin wave dispersion for this material.
The data reveal rather isotropic spin wave cones, {\em i.e.} that $v_{\perp}$ and $v_{\parallel}$ are rather similar, with
$v_{\perp} \approx (1.03 \pm 0.06) v_{AF}$ and $v_{\parallel} \approx (0.86 \pm 0.06) 
v_{AF}$,\cite{john}  where $v_{AF}$ is the acoustic spin wave velocity of the undoped antiferromagnet.
For the DS3 state, if we include {\em only} $J_b$ or {\em only} $J_c$, we find no coupling 
strength $\lambda$ for which these two relations can be simultaneously satisifed.
The presence of the two couplings together, as shown in Fig.~\ref{fig:vel.ds} with $J_b = J_c$,
can account for the proper relationship among the velocities, but only for 
a small range of rather small coupling ratio.
As a general trend, we find approximate isotropy of the spin wave cones
to be more robust for bond-centered stripes
(in both vertical and diagonal cases), 
and so one might suspect bond-centered stripes could be responsible for the
near isotropy of the spin waves in this material.
However, as we have shown, the DB3 configuration yields a 
ferromagnetic spin wave dispersion, which is certainly not supported by the data. 

In the related compound La$_{2}$NiO$_{4.133}$\cite{sitecentered}, 
signatures of spin stripes have been detected in neutron scattering.  
``Incommensurate peaks''
are observed to persist up to a temperature $T_{m}$, above which 
magnetic peaks indicative of stripe structure can be regained by application of a 
$6T$ magnetic field.  The field-induced stripe spacing (both above and slightly below $T_{m}$) 
is smaller than the zero-field stripe spacing observed below $T_{m}$. 
As noted by the authors,\cite{sitecentered} the ferrimagnetic
response is naturally explained by bond-centered stripes.
In the high temperature field-induced stripe phase, the diagonal stripes 
have spacing $p=3$.  
Our results in Fig.~\ref{dbond}
suggest that this field-induced transition should be accompanied by a 
dramatic change in the low-energy spin wave dispersion, from linear to quadratic.

We have also shown that (as in the site-centered case\cite{kruger}) the number of bands in a bond-centered
configuration is set by the number of 
spins in the unit cell, rather than by the spacing $p$.  
Generally, for both vertical and diagonal stripes, 
site-centered stripes have $(p-1)$ spin wave bands, and bond-centered stripes have $p$ bands.  The exception is
the case of diagonal site-centered stripes with odd spacing $p$, which has $\frac{1}{2}(p-1)$ bands.  
An experimental consequence of this is that for a given value of $p$, bond-centered stripes have 
$p$ spin wave bands, whereas site-centered stripes have at most $p-1$ bands.  Although not yet observed experimentally,
this means that the upper bands can also be used to distinguish site- from bond-centered stripes.  Finding $p$ bands along with
incommensurate peaks indicative of spacing $p$ would rule out site-centered stripes.  
For diagonal odd width stripes, the threshold is even lower.  For, {\em e.g.,} $DS3$, 
only one spin wave band is expected, whereas for DB3, we expect to find three bands.
The observance of a second band (or eqivalently a spin wave crossing) for diagonal $p=3$
sripes would rule out a site-centered configuration.
Of course, negative evidence is dicier, and 
the observance of the smaller number of bands cannot distinguish the two, as it cannot rule out the possibility
that the top band is too faint to be observed.  

\section{Conclusions}

In conclusion, we have studied regular arrays of antiphase domain walls in
two-dimensional Heisenberg antiferromagnets and find that 
their location relative to the lattice--{\em i.e.}, 
whether they are site-centered or
bond-centered--produces distinct effects which may be measurable in a diffraction
probe such as neutron scattering.  
In particular, arrays of odd-width, bond-centered
antiphase domain walls generically produce more elastic peaks than 
site-centered stripes. In addition, bond-centered stripes generically produce more bands than site-centered stripes.
We further find that low-energy
spin wave velocities are not always directly related to the exchange
couplings in the model, and in particular for bond-centered configurations,
rather isotropic spin wave cones are predicted for a wide range of parameters.

\section*{Acknowledgements}
It is a pleasure to thank I.~Affleck, A.~Castro Neto, S.~Kivelson,
B.~Lake, Y.~S.~Lee, S.~Rosenkranz, A.~Sandvik, and J.~Tranquada for helpful discussions.  
This work was supported by the U.S. Government and by Boston University.

\appendix

\section{Spin-wave methods \label{hp}}

We rewrite the Hamiltonian Eqn.~(\ref{model}) using the ladder
operators:
\begin{equation}
H=\frac{1}{2} \sum_{<\mathbf{r},\mathbf{r'}>} J_{\mathbf{r},\mathbf{r'}}
[S^z_{\mathbf{r}}S^z_{\mathbf{r'}} + \frac{1}{2}(S^+_{\mathbf{r}}S^-_{\mathbf{r'}}+S^-_{\mathbf{r}}S^+_{\mathbf{r'}})].
\end{equation}
We now replace the spin
operators by Holstein-Primakoff (HP) bosons\cite{auerbach} 
\begin{eqnarray}
S_i^+= && \sqrt{2 S} a_i \nonumber \\
S_i^-= && \sqrt{2 S} a_i^+ \nonumber \\
S_i^z=&&S-a_i^+ a_i
\end{eqnarray}
for odd sites $i$ occupied by a spin up, and 
\begin{eqnarray}
S_i^+= && \sqrt{2 S} a_i^+ \nonumber \\
S_i^-= && \sqrt{2 S} a_i \nonumber \\
S_i^z= &&-S + a_i^+ a_i
\end{eqnarray}
for even sites $i$ occupied by a spin down. 
Here, $i$ labels each spin within a unit cell,
i.e. $i=1,2,\cdots,N$, where $N$ is the number of spins in the unit cell. We use
odd $i$ to represent $S_z=\uparrow$ spins and even i for  $S_z=\downarrow$ spins.
We Fourier transform the bosonic operators via
\begin{eqnarray}
a_i (\mathbf{k})=&&\frac{1}{\sqrt{n}} \sum_{\mathbf{r}\in odd ~i} a_{\mathbf{r}}
e^{i {\mathbf{k}} \cdot {\mathbf{r}}}, \nonumber \\
a_j (\mathbf{k})=&&\frac{1}{\sqrt{n}} \sum_{\mathbf{r}\in even ~i} a_{\mathbf{r}}
e^{-i {\mathbf{k}} \cdot {\mathbf{r}}}.
\end{eqnarray}
Finally, we get the Hamiltonian in momentum space
\begin{eqnarray}
H&=&\sum_{ij} A_{i,j} a_i^+(\mathbf{k}) a_j(\mathbf{k})  \nonumber \\
&+&\frac{1}{2}\sum_{ij} [B_{ij} a_i^+(\mathbf{k}) a_j^+(\mathbf{k}) + B_{ij}^*a_j(\mathbf{k})
a_i(\mathbf{k})], \label{quadh}
\end{eqnarray}
where $A=A^+$ and $B^T=B$.

The quadratic Hamiltonian (\ref{quadh}) can be diagonized 
via a canonical symplectic transformation\cite{blaizot} $T$, $b = T a$, using the bosonic metric
\begin{equation}
\eta=\left( \begin{array}{ccc}
I & 0 \\
0  &  -I \\
\end{array}\right)~,
\end{equation}
where $I$ is the $N \times N$ indentity matrix.
This leads to 
\begin{equation}
H(\mathbf{k})= \sum_{\alpha} [b^+_{\alpha}(\mathbf{k}) \omega_{\alpha} (\mathbf{k})
b_{\alpha}(\mathbf{k}) + \frac{1}{2} \omega_{\alpha} (\mathbf{k})].
\end{equation}

We now consider the structure factor.  Only 
$S^x$ and $S^y$ 
contribute to the inelastic part of the structure factor. 
In terms of HP bosons, 
\begin{eqnarray}  \label{sxa}
S_{\mathbf{k}}^x&=&\frac{1}{2} (S_{\mathbf{k}}^+ +S_{\mathbf{k}}^-) 
\\
&=&\sqrt{\frac{S}{2}}  \bigg( \sum_{i \in odd} [a_i^+(-\mathbf{k}) + a_i
({\mathbf{k}})] \nonumber \\
&&+ \sum_{i \in even} [a_i^+(\mathbf{k}) + a_i
({-\mathbf{k}})] \bigg)  \nonumber 
\end{eqnarray}

We then substitute Eqn.~(\ref{sxa}) into the structure factor and  
keep only the creation operators $\{b_1^+
  (\mathbf{k}),b_2^+
  (\mathbf{k}),\cdots \}$, which connect the vacuum to singly excited states. This gives
\begin{eqnarray}
S^{in} (\mathbf{k}, \omega_{\alpha})=&&2 \sum_f |<f|S^x_{\mathbf{k}}|~0>|^2 \delta
(\omega-\omega_f) \nonumber \\
=&&S |<1~|(\sum_{i} ~\alpha_{i} b_i^+ )|~0>|^2  \nonumber \\
=&&S |\sum_{i}~\alpha_{i}|^2,
\end{eqnarray}
where $\alpha_i$ is the ith component of 
the (orthonormalized) eigenvector $|\alpha>$ of the Hamiltonian using the bosonic metric,
corresponding to eigenvalue $\omega_{\alpha}$.

\section{Dimerized spin model \label{dimer}}

We consider an isolated system of two spins with ferromagnetic coupling $J_b$. In the ground state,
the two spins are aligned. When the spins tilt a bit, each produces an
effective field acting on the other. Using the classical spin wave method,
we have
\begin{eqnarray}
\frac{d \mathbf{S}_1 }{d t}= && -\frac{J_b}{\hbar} \mathbf{S}_1 \times
\mathbf{S}_2 \nonumber \\
\frac{d \mathbf{S}_2 }{d t}= && -\frac{J_b}{\hbar} \mathbf{S}_2 \times
\mathbf{S}_1 
\end{eqnarray}
Ignoring the change in $S_z$, the x, y components of the two spins satisfy
\begin{eqnarray}
\frac{d S_1^x }{d t} = -\frac{d S_2^x }{d t}&=&- \frac{J_b S}{\hbar} (S_1^y-S_2^y) \nonumber \\
\frac{d S_1^y }{d t} = -\frac{d S_2^y }{d t}&=&\frac{J_b S}{\hbar} (S_1^x-S_2^x)~. \nonumber \\
\end{eqnarray}
Integrating yields $S_1^{x,y}=-S_2^{x,y}+ c$, where $c$ is a constant of integration. 
Since we allow only $S_z$ to have a constant component, $c=0$.  
Taking the second derivative of 
$S_1^x$, we find
\begin{eqnarray}
\frac{d^2 S_1^x}{d t^2}=&&-\frac{J_b S}{\hbar} (\frac{d S_1^y}{d t}- \frac{d
  S_2^y}{d t}) \nonumber \\
=&&-\frac{J_b S}{\hbar^2}  [J_b S (S_1^x- S_2^x) +J_b S (S_1^x- S_2^x)] \nonumber \\
=&&-2 \frac{J_b^2 S^2}{\hbar^2} [S_1^x- S_2^x] \nonumber \\
=&&-4 \frac{J_b^2  S^2}{\hbar^2} S_1^x~,
\end{eqnarray}
which is a harmonic oscillator equation. If we set $S_1^x (x,t)= u(x)
e^{i \omega t}$, we see that the oscillation frequency is
\begin{equation}
\omega= 2 \frac{|J_b|}{\hbar} S,
\end{equation}
which recovers the large $J_b$ limit of Eqn.~\ref{eqn:b2omega}.


\begin{thebibliography}{10}
\bibitem{foot1}
{Here we follow the convention of the literature, 
and express the wavevector in units where the lattice spacing $a=1$.}
\bibitem{foot2}
{We follow the convention of the literature,
and refer to satellite peaks around $(\pi,\pi)$ as incommensurate peaks, even though
they may be commensurate with the underlying lattice.}
\bibitem{foot3}
{In the physical system, holes residing on the domain
wall do carry spin, and they may under certain conditions lead to a net
spin on the domain walls as well.  However, this is a secondary order parameter
which generically can order only after the spin texture we describe, and
we leave its description to future work.}
\end{thebibliography}
\bibliographystyle{forprb}

\end{document}